\documentclass[floatfix,amsmath,amssymb,aps,prc,twocolumn,superscriptaddress,showpacs,preprintnumbers,nofootinbib]{revtex4-1}
\usepackage{graphicx}
\usepackage{dsfont}
\usepackage{xcolor}
\usepackage{multirow}
\usepackage[normalem]{ulem}
\usepackage{enumitem} 
\usepackage{lineno}
\usepackage{bm}
\usepackage{charter} 
\usepackage{subfigure} 
\definecolor{refcol}{RGB}{34,34,178}
\usepackage{nameref}
\usepackage[colorlinks,linkcolor=blue,citecolor=blue,urlcolor=blue]{hyperref}
\usepackage{microtype}
\usepackage{float}
\usepackage{appendix}
\usepackage{multirow}
\usepackage{soul}
\usepackage{booktabs} 
\usepackage{subfigure}  
\usepackage[utf8]{inputenc}
\usepackage{textgreek}
\usepackage{amsmath}

\usepackage{graphicx,color}
\usepackage{multirow}
\usepackage{times}
\usepackage{bm}
\usepackage{epstopdf}
\usepackage{enumerate}
\usepackage{ulem}

\graphicspath{{./Fig/}}
\begin{document}
	
	\title{
		Resolving Ratio Redundancy in Chemical Freeze-out Studies with Principal Component Analysis and Bayesian Calibration
	}      
	
	\author{Nachiketa Sarkar}
	\email{nachiketa.sarkar@gmail.com}
	\affiliation{Department of Physics, Amrita School of Physical Sciences, Coimbatore, Amrita Vishwa Vidyapeetham, India} 
	
	\date{\today}
\begin{abstract}
	We introduce a Principal Component Analysis (PCA)--Bayesian framework for extracting chemical freeze-out conditions in relativistic heavy-ion collisions that resolves long-standing ambiguities in hadron-ratio--based analyses. By constructing all possible hadron-yield ratios from a chosen set of species and transforming them into an orthogonal PCA basis, the method removes linear redundancies and eliminates the information loss and systematic uncertainties associated with ratio selection. Energy-wise Bayesian calibration of the Hadron Resonance Gas (HRG) model is then performed directly in this decorrelated space, with a Gaussian Process emulator enabling fast and accurate model evaluations. A detailed Sobol sensitivity analysis, together with the PCA loading structure, identifies the most informative ratio combinations and reveals a transition from chemical-potential--dominated to temperature-controlled freeze-out with increasing $\sqrt{s_{NN}}$. The calibrated model reproduces all measured ratios, and the extracted freeze-out parameters are consistent with previous HRG determinations.
\end{abstract}

	\maketitle

  \section{Introduction}
   
  The quark–gluon plasma (QGP)~\cite{Shuryak:1980tp}, an emergent state of matter formed in ultra-relativistic heavy-ion collisions, undergoes rapid expansion and cooling as it evolves from a deconfined medium toward hadronization. As the system cools, inelastic interactions cease and the relative abundances of hadron species become fixed. This chemical freeze-out stage is characterized by the temperature $T_{\rm ch}$ and the baryon, strangeness, and electric-charge chemical potentials $(\mu_B,\mu_S,\mu_Q)$~\cite{Braun-Munzinger:1994ewq,Cleymans:1999st}. These thermodynamic parameters quantitatively relate the observed hadron yields to the underlying properties of strongly interacting matter. Their systematic dependence on collision energy—studied across AGS, SPS, RHIC, and LHC programs—provides essential information on the Quantum Chromodynamics (QCD) phase structure and the conditions relevant for deconfinement and potential critical phenomena~\cite{Stephanov:1998dy,Stephanov:1999zu,Stephanov:2004wx,Tawfik:2004sw}.

 Thermal statistical models, in particular the Hadron Resonance Gas (HRG) model, have been widely employed to extract these freeze-out parameters~\cite{Braun-Munzinger:1994ewq,Cleymans:1996cd,Cleymans:1998fq,Cleymans:1999st,Braun-Munzinger:1999hun,Braun-Munzinger:2001hwo,Becattini:2003wp,Braun-Munzinger:2003pwq,Andronic:2005yp,Becattini:2005xt,Andronic:2008gu,Andronic:2011yq,Begun:2012rf,Andronic:2017pug,Chatterjee:2017yhp,Alba:2014eba,Adak:2016jtk,Cleymans:2005xv,Chatterjee:2013yga,Vovchenko:2015cbk,Vovchenko:2016ebv,Alba:2016hwx,Alba:2017bbr,Sarkar:2025bkc,Yen:1997rv,Tiwari:2011km}. Numerous refinements have been developed, including excluded-volume interactions~\cite{Rischke:1991ke,Cleymans:1992jz,Yen:1997rv,Tiwari:2011km,Begun:2012rf,Andronic:2012ut,Bhattacharyya:2013oya,Vovchenko:2016rkn,Sarkar:2018mbk,Alba:2016hwx,Alba:2017bbr,Sarkar:2025bkc}, extended hadron lists~\cite{Vovchenko:2014pka,Alba:2017bbr,Sarkar:2017ijd,Karthein:2021cmb,Noronha-Hostler:2012ycm,Sarkar:2017bqy}, finite-size effects~\cite{Albright:2014gva,Sarkar:2017ijd,Sarkar:2019oyo}, treatments based on scattering phase shifts~\cite{Venugopalan:1992hy,Dash:2018can,Vovchenko:2017drx,Dash:2018mep}, and approaches incorporating a strangeness-suppression factor $\gamma_s$~\cite{Koch:1986ud,Becattini:2003wp,Manninen:2008mg,Cleymans:2005xv,Biswas:2020dsc} to model incomplete equilibration of strange quarks. Alternative freeze-out scenarios, such as flavor-dependent or sequential freeze-out~\cite{Bugaev:2013sfa,Chatterjee:2013yga,Chatterjee:2014ysa,Flor:2020fdw}, and system-size-dependent schemes~\cite{Sharma:2018jqf,Panda:2021zab,Flor:2021olm}, have also been explored to accommodate potential species-dependent decoupling.
 
 Despite these various statistical-model refinements and methodological approaches, traditional extractions of freeze-out parameters still rely primarily on $\chi^{2}$ minimization of experimental hadron multiplicities or their ratios. Ratio-based fits, however, face two structural limitations. First, for $N$ measured yields, forming the full set of $\sim N(N-1)/2$ ratios introduces redundancy because only $N-1$ ratios are linearly independent~\cite{Andronic:2005yp,Bhattacharyya:2019cer,Bhattacharyya:2020sgn,Manninen:2008mg}. Second, restricting the analysis to a subset of ratios avoids over-counting but discards part of the information and may bias the extracted parameters~\cite{Andronic:2005yp}. However, as emphasized by Becattini~\cite{Becattini:2007wt}, constructing ratios \emph{a posteriori} from measured yields and treating them as independent observables violates the assumptions of least-squares fitting, as it neglects correlations arising from shared denominators and common systematic uncertainties. This omission renders the resulting $\chi^{2}$ statistically inconsistent and can bias the inferred freeze-out parameters~\cite{Manninen:2008mg}. Although directly measured ratios (e.g., $\pi^{-}/\pi^{+}$ or $p/\bar{p}$) avoid this issue, they still do not resolve the information loss inherent in fitting fewer than the full set of independent ratios. Direct yield fits, on the other hand, avoid ratio redundancy but introduce the system volume as an additional fit parameter and often suffer from parameter degeneracies between the volume and the thermodynamic variables~\cite{Torrieri:2004zz,Sarkar:2025bkc}.

Several studies have attempted to address these difficulties. Bhattacharyya \textit{et al.}~\cite{Bhattacharyya:2019cer} analyzed the uncertainties introduced by ratio selection, while Ref.~\cite{Bhattacharyya:2020sgn} proposed sampling statistically independent spanning trees of ratios to quantify this effect. Another method, based on four global constraints, was introduced in Ref.~\cite{Bhattacharyya:2019wag}; however, the required choice of specific ratios and the sensitivity to feed-down corrections and acceptance mismatches introduce additional systematic uncertainties. Furthermore, the imposition of exactly four constraints—and the fact that their selection is not unique—adds an additional source of ambiguity in the extracted freeze-out parameters.

These considerations motivate the development of an approach that retains the full information content of the measured yields while avoiding ambiguities associated with ratio selection and correlated observables. In this work, we adopt a data-driven strategy in which all possible hadron ratios are constructed and then transformed using Principal Component Analysis (PCA)~\cite{Jolliffe2016,Pearson1901}. The resulting PCA basis provides an orthogonal set of statistically independent combinations of ratios, removing linear correlations and redundancy while preserving the full set of independent degrees of freedom. This procedure directly addresses the correlation issues highlighted by Becattini~\cite{Becattini:2007wt}: correlations among hadron ratios are not ignored but are systematically treated through orthogonalization, ensuring that each transformed observable carries a distinct statistical weight. The Bayesian calibration of the freeze-out parameters is then performed using these principal components rather than the raw ratios, allowing each independent direction in the data to contribute consistently and avoiding the information loss or over-counting that can arise in conventional $\chi^{2}$ analyses.

 To isolate the intrinsic energy dependence of chemical freeze-out and minimize complications arising from centrality-related correlations, we restrict our analysis to the most central collisions, covering a beam energy range from $\sqrt{s_{NN}} = 7.7~\mathrm{GeV}$ (RHIC Beam Energy Scan) up to $2.76~\mathrm{TeV}$ (LHC). For each energy, we perform two independent fits—with and without explicit conserved-charge constraints—to assess their influence on the extracted parameters. The methodology developed here provides a consistent framework for determining freeze-out conditions and can be extended in future work to incorporate centrality dependence and a more detailed treatment of systematic uncertainties.

 The remainder of this paper is organized as follows. Section~\ref{sec:HRG} presents a brief overview of the Hadron Resonance Gas (HRG) model. Section~\ref{sec:analysis_methodology} describes the analysis methodology. Section~\ref{sec:results_discussion} presents the Bayesian calibration results together with a detailed discussion. Finally, Section~\ref{sec:conclusion} summarizes the main conclusions of this work.

\section{Hadron Resonance Gas (HRG) Model}
\label{sec:HRG}

The Hadron Resonance Gas (HRG) model provides a widely used and effective description of hadronic matter in the confined phase of QCD~\cite{Braun-Munzinger:1994ewq,Cleymans:1996cd,Cleymans:1998fq,Cleymans:1999st,Braun-Munzinger:1999hun,Braun-Munzinger:2001hwo,Becattini:2003wp,Braun-Munzinger:2003pwq,Andronic:2005yp,Becattini:2005xt,Andronic:2008gu,Andronic:2011yq,Begun:2012rf,Andronic:2017pug,Chatterjee:2017yhp,Alba:2014eba,Adak:2016jtk,Cleymans:2005xv,Chatterjee:2013yga,Vovchenko:2015cbk,Vovchenko:2016ebv,Alba:2016hwx,Alba:2017bbr,Sarkar:2025bkc,Yen:1997rv,Tiwari:2011km,Rischke:1991ke,Cleymans:1992jz,Andronic:2012ut,Bhattacharyya:2013oya,Vovchenko:2016rkn,Sarkar:2018mbk,Vovchenko:2014pka,Sarkar:2017ijd,Karthein:2021cmb,Noronha-Hostler:2012ycm,Sarkar:2017bqy,Venugopalan:1992hy,Dash:2018can,Vovchenko:2017drx,Dash:2018mep,Karsch:2003zq,Koch:1986ud,Manninen:2008mg,Bugaev:2013sfa,Chatterjee:2014ysa,Flor:2020fdw,Sharma:2018jqf,Panda:2021zab,Flor:2021olm,Bhattacharyya:2019cer,Bhattacharyya:2020sgn,Tawfik:2004sw,Garg:2013ata,Albright:2014gva,Sarkar:2019oyo,Torrieri:2004zz,Bhattacharyya:2019wag,ParticleDataGroup:2020ssz}. In the ideal HRG formulation, the hadronic medium is described as a non-interacting gas of all known hadrons and resonances, each contributing to the thermodynamic properties according to its mass, spin, and conserved quantum numbers. For a hadron species $i$ with mass $m_i$, degeneracy $g_i$, and chemical potential $\mu_i$, the equilibrium momentum distribution is given by the Bose--Einstein or Fermi--Dirac form,
\begin{equation}
	f_i(E_i) = \frac{1}{\exp\!\left[\frac{E_i - \mu_i}{T_\mathrm{ch}}\right] \pm 1},
\end{equation}
where the $+$ ($-$) sign corresponds to fermions (bosons), $E_i = \sqrt{p^2 + m_i^2}$ is the single-particle energy, and $T_\mathrm{ch}$ denotes the chemical freeze-out temperature.

The primordial number density of species $i$ follows from the momentum integral,
\begin{equation}
	n_i(T_\mathrm{ch},\mu_i)
	= \frac{g_i}{2\pi^2}
	\int_0^\infty 
	\frac{p^2\, dp}{
		\exp\!\left[\frac{\sqrt{p^2 + m_i^2} - \mu_i}{T_\mathrm{ch}}\right] \pm 1 }.
\end{equation}

To incorporate resonance decays, the final observable yield of species $i$ is computed as
\begin{equation}
	N_i 
	= V_{\text{sys}}
	\!\left[
	n_i(T_\mathrm{ch},\mu_i)
	+ \sum_j n_j(T_\mathrm{ch},\mu_j)\, B_{j\rightarrow i}
	\right],
\end{equation}
where $B_{j\rightarrow i}$ denotes the branching ratio for the decay of resonance $j$ into hadron $i$, and $V_{\text{sys}}$ is the effective system volume. For yield ratios, the volume factor cancels, leaving the ratios dependent only on the thermal parameters $(T_\mathrm{ch},\mu_B,\mu_S,\mu_Q)$.

In this analysis, the HRG calculation employs all established hadrons and resonances with masses up to $3~\mathrm{GeV}$, following the latest Particle Data Group (PDG) compilation~\cite{ParticleDataGroup:2020ssz}.

\section{Analysis Methodology}
\label{sec:analysis_methodology}

\subsection{Construction of Experimental Ratios and Covariance}
 
The analysis begins with the $N$ experimentally measured hadron yields and their reported statistical uncertainties,
\begin{equation}
	\mathbf{y} = (y_1, y_2, \ldots, y_N),  
	\qquad 
	\boldsymbol{\sigma} = (\sigma_1, \sigma_2, \ldots, \sigma_N).
\end{equation}

From these yields, all distinct pairwise ratios are constructed as
\begin{equation}
	r_{ij} = \frac{y_i}{y_j}, 
	\qquad 
	1 \le i < j \le N,
\end{equation}
forming the ratio vector
\begin{equation}
	\mathbf{r} = (r_1, r_2, \ldots, r_M), 
	\qquad 
	M = \frac{N(N-1)}{2}.
\end{equation}

To propagate the experimental uncertainties to the ratios, Monte Carlo (MC) sampling of the yields is performed.  In the absence of published experimental covariance matrices, the yield uncertainties are treated as independent Gaussian fluctuations*,
\begin{equation}
	\begin{aligned}
		\mathbf{y}^{(k)} &\sim \mathcal{N}(\mathbf{y}, \Sigma_y), 
		\qquad
		\Sigma_y = \mathrm{diag}(\sigma_1^2, \ldots, \sigma_N^2), \\
		k &= 1, \ldots, N_\mathrm{MC}.
	\end{aligned}
\end{equation}
 
For each sampled yield vector, the corresponding ratios are computed as
\begin{equation}
	r_{ij}^{(k)} = \frac{y_i^{(k)}}{y_j^{(k)}}, 
	\qquad 
	\mathbf{r}^{(k)} = \big(r_1^{(k)}, \dots, r_M^{(k)}\big).
\end{equation}

The complete ensemble of sampled ratios forms the matrix
\begin{equation}
	R \in \mathbb{R}^{N_\mathrm{MC} \times M},
\end{equation}
 
where each row corresponds to a single Monte Carlo realization.  
A sample size of $N_\mathrm{MC} = 5000$ was found sufficient, as doubling the sample size modifies the covariance eigenvalues by less than $0.1\%$.

From this ensemble, the mean ratio vector and covariance matrix are computed as
\begin{align*}
	\bar{\mathbf{r}} &= \frac{1}{N_\mathrm{MC}} 
	\sum_{k=1}^{N_\mathrm{MC}} \mathbf{r}^{(k)}, \\
	C_r &= \frac{1}{N_\mathrm{MC}-1} 
	\sum_{k=1}^{N_\mathrm{MC}} 
	\left(\mathbf{r}^{(k)} - \bar{\mathbf{r}}\right)
	\left(\mathbf{r}^{(k)} - \bar{\mathbf{r}}\right)^{T}.
\end{align*}
 
 This MC construction preserves the statistical correlations generated by the ratio definition itself, since each sampled yield vector contributes coherently to all ratios. Ratios sharing common yields therefore fluctuate together across the ensemble, and these induced dependencies appear in the off-diagonal elements of the covariance matrix $C_r$.
 Full experimental covariance matrices for identified hadron yields are not publicly available, so possible cross-species systematic correlations cannot be included explicitly. The assumption of independent Gaussian uncertainties is therefore applied at the level of the yields. The ratio construction nevertheless captures the dominant correlation structure arising from shared yields. While neglecting unreported experimental covariances may slightly affect the relative weighting of the principal components, it does not alter their orientation or the physical interpretation of the PCA basis. Within these limitations, the PCA transformation remains sufficiently stable for the subsequent analysis.

\subsection{PCA Analysis}
 
\begin{figure} 
	\centering
	\includegraphics[scale=0.3]{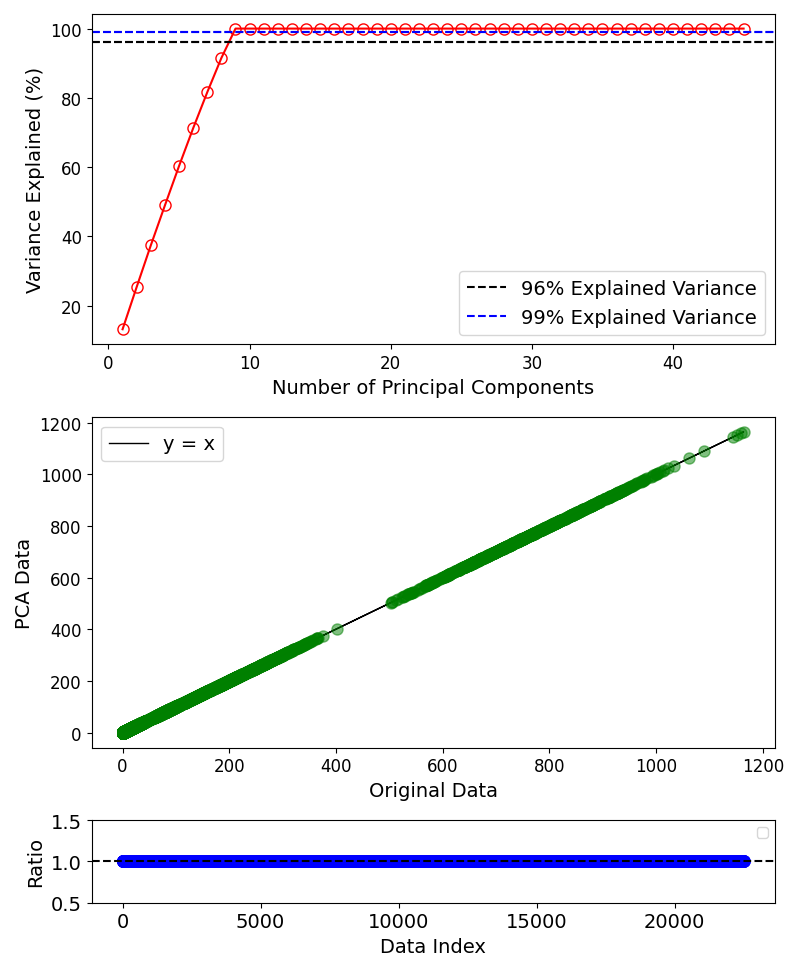}
	\caption{
		PCA diagnostics for the hadron yield ratio ensemble. 
		\textbf{Top:} Cumulative variance explained as a function of the number of principal components, with dashed lines marking the 96\% and 99\% thresholds. 
		\textbf{Middle:} Comparison between the standardized log-ratio data and the reconstruction from the retained components. 
		\textbf{Bottom:} Ratio of the reconstructed to the original data.
	}
	\label{fig:pca_diagnostics_plot}
\end{figure}

PCA is used to transform the correlated hadron ratios into an orthogonal set of observables. This decorrelates the ratio space and restricts the analysis to the independent degrees of freedom relevant for Bayesian calibration. However, Before performing PCA, we work with \emph{log-ratios}, which convert the nonlinear redundancy among raw ratios into linear constraints. For two species $i$ and $j$ the log-ratio is
\begin{equation}
	R_{ij} = \ln y_i - \ln y_j.
\end{equation}
With $N$ species, only $N-1$ independent combinations exist. In log space these constraints are linear, allowing PCA to recover exactly $N-1$ nonzero components.

Let $\mathcal{R}$ denote the log-ratio ensemble matrix, with each row corresponding to a MC sample and each column to a distinct log-ratio. Before PCA, each column of $\mathcal{R}$ is standardized to have zero mean and unit variance. The eigen-decomposition of $\mathcal{R}^T \mathcal{R}$ is then performed,
\begin{equation}
	\mathcal{R}^T \mathcal{R} = V \, \Sigma^2 \, V^T ,
\end{equation}
where $\Sigma$ contains the singular values and $V$ defines the orthogonal transformation.

The principal components are obtained by projecting the log-ratio ensemble onto these eigenvectors:
\begin{equation}
	Z = \mathcal{R} V.
\end{equation}
Because only $N-1$ independent combinations exist, we retain exactly $q=N-1$ principal components,
\begin{equation}
	Z_q = \mathcal{R} V_q,
\end{equation}
thereby preserving all independent information.  As also evident from the top panel of Fig.~\ref{fig:pca_diagnostics_plot}, the cumulative explained variance saturates after the ninth component. The comparison between the original and PCA-reconstructed log ratios (middle and bottom panels) indicates that these nine components retain all the (as all higher components have vanishing eigenvalues) information needed to describe the ensemble.

\begin{figure} 
	\centering
	\includegraphics[scale=0.5]{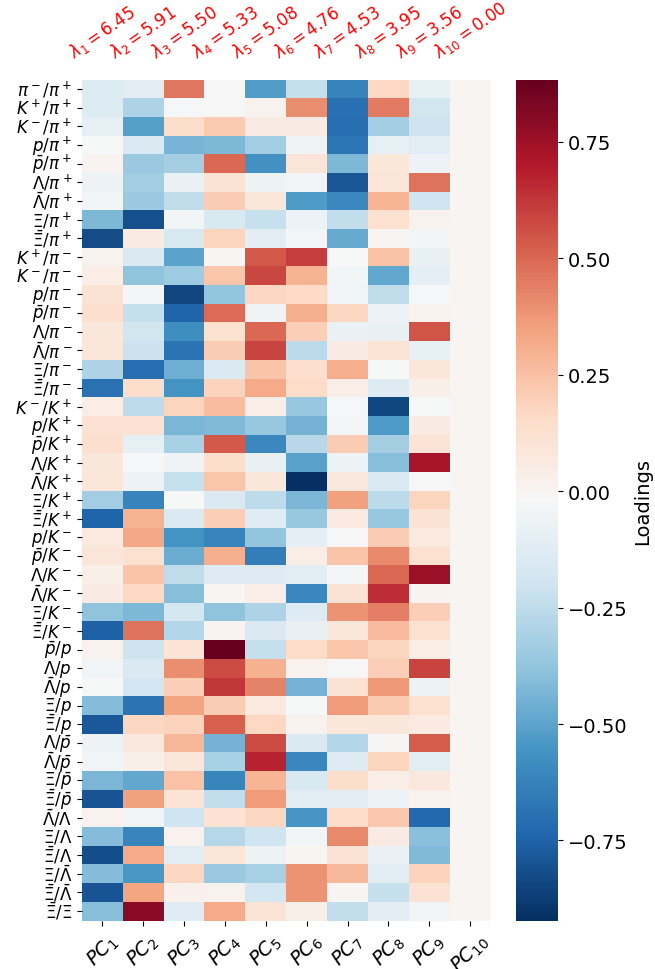}
	\caption{
		Loading matrix representing the first ten principal components (PC1--PC10) for identified-particle yield log-ratios at $\sqrt{s_{NN}}=39$~GeV. 
		The eigenvalues $\lambda_i$ are indicated for each component. 
		The color bar shows the contribution of each log-ratio to each principal component.
	}
	\label{fig:loading_7.7}
\end{figure}

In the PCA framework, the product $V\Sigma$ (up to normalization conventions) is referred to as the loading matrix. Its columns correspond to the principal components, and its rows to the original log-ratios. Each entry quantifies the direction and variance-weighted contribution of a given ratio to a component. The loading matrix therefore offers physical insight, as it helps identify the dominant sources of correlated fluctuations within the ratio ensemble. Figure~\ref{fig:loading_7.7} shows the loading matrix at $\sqrt{s_{NN}} = 39~\mathrm{GeV}$. Only the first nine components exhibit nonzero eigenvalues, consistent with the expected $N{-}1$ independent degrees of freedom. The eigenvalue pattern illustrates that PCA removes the near-singular directions of the ratio covariance matrix, reducing its condition number from $\kappa=\lambda_{\max}/\lambda_{\min} \sim \infty$ to $\mathcal{O}(1)$ for the nine retained components. This indicates that the transformed basis is numerically well-conditioned for subsequent Bayesian inference. A detailed discussion of the physical interpretation of the energy-dependent loading patterns is provided in Sec.~\ref{sec:results_discussion}.


\subsection{Gaussian Process Regression and Construction of Model Input}

  Markov Chain Monte Carlo (MCMC) sampling during Bayesian inference requires repeated evaluation of model predictions across a large number of parameter points. Performing full HRG calculations at every step would be computationally expensive. To reduce this cost, we construct a Gaussian Process (GP) emulator that provides interpolated model predictions together with their associated uncertainties.
   
  A set of 500 design points in the four-dimensional parameter space $(T, \mu_B, \mu_S, \mu_Q)$ is generated using Latin Hypercube Sampling (LHS)~\cite{McKay1979LHS}. At each point, the full HRG model—including all resonance decay contributions—is evaluated to obtain the corresponding hadron densities. Of these, 450 points are used for GP training and 50 points are retained for validation.

\begin{figure} 
	\centering
	\includegraphics[scale=0.30]{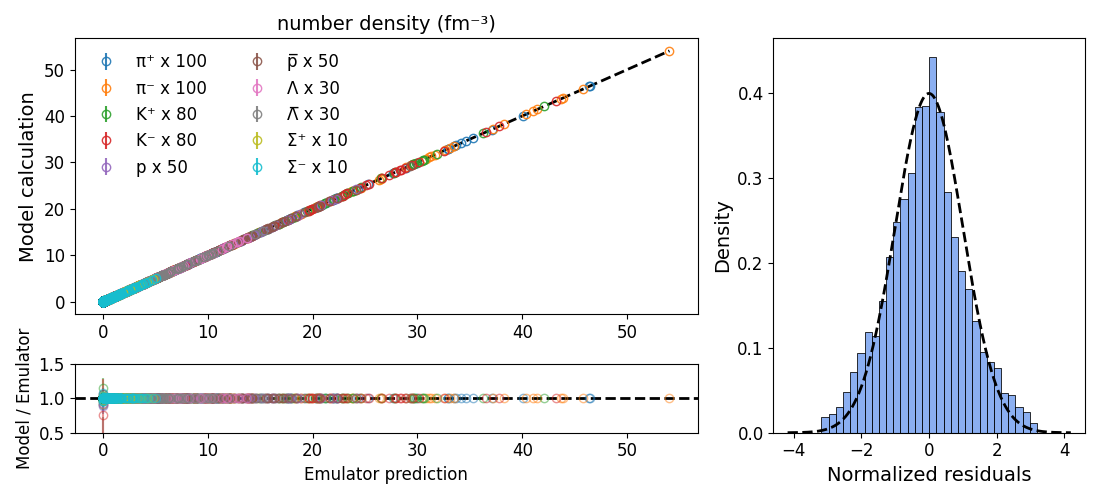}
	\caption{
		GP Emulator diagnostics. 
		\textbf{Top-left:} HRG model calculations versus emulator predictions for various hadron species , with error bars scaled to the predicted uncertainty. 
		The dashed line denotes perfect agreement. 
		\textbf{Bottom-left:} Ratio of model to emulator predictions, showing deviations tightly clustered around unity. 
		\textbf{Right:} Distribution of normalized residuals with a standard normal overlay, demonstrating statistically consistent uncertainty estimates.
	}
	\label{fig:GPR_validation}
\end{figure}
   
  The GP emulator employs a squared-exponential (radial basis function) kernel,
  \begin{equation}
  	k(x, x') = \sigma^2_{\mathrm{GP}}
  	\exp \left( - \sum_k \frac{(x_k - x_k')^2}{2 l_k^2} \right)
  	+ \sigma^2_n \, \delta_{x x'} ,
  \end{equation}
  where $\sigma_{\mathrm{GP}}^{2}$ is the signal variance, $l_k$ are the characteristic length scales, and $\sigma_n^{2}$ is a small diagonal term included for numerical stability. The hyperparameters $(\sigma_{\mathrm{GP}}, l_k, \sigma_n)$ are determined by maximizing the marginal likelihood.

 The emulator performance is assessed using the 50 validation points, as shown in Fig.~\ref{fig:GPR_validation}. Emulator predictions follow the HRG model values closely over the sampled range. The ratio plot (bottom-left) shows residuals centered around unity, and the normalized residuals (right panel) are approximately standard normal, indicating that both the mean predictions and the associated uncertainty estimates are consistent with the validation data.
 
 \medskip
		
\medskip
Starting from the emulator-predicted mean yields $\boldsymbol{\mu}_m$, we construct the corresponding set of model log ratios using the same index pairs $(i_k, j_k)$ as in the experimental analysis:
\begin{equation}
	\mathcal{R}^{\mathrm{model}}_{i,j} = \log\!\left(\frac{\mu_i}{\mu_j}\right),
	\qquad 1 \le i < j \le N.
\end{equation}

\bigskip		
textbf{Covariance propagation in ratio space.}
To quantify emulator-induced uncertainties in log-ratio space, we propagate the yield covariance matrix using linear error propagation. For
\[
\mathcal{R}_{i,j}^{\mathrm{model}} = \ln(\mu_i) - \ln(\mu_j),
\]
the Jacobian elements are
\begin{equation}
	\frac{\partial \mathcal{R}_{i,j}^{\mathrm{model}}}{\partial \mu_\ell} =
	\begin{cases}
		+\dfrac{1}{\mu_i}, & \ell = i, \\[2mm]
		-\dfrac{1}{\mu_j}, & \ell = j, \\[2mm]
		0, & \text{otherwise.}
	\end{cases}
\end{equation}
This yields the Jacobian matrix $J \in \mathbb{R}^{N_R \times N}$, where $N_R = N - 1$ is the number of independent log ratios. The ratio-space covariance matrix is then
\begin{equation}
	C_R^{\mathrm{model}} = J \, C_y^{\mathrm{GPR}} \, J^\top ,
\end{equation}
with $C_y^{\mathrm{GPR}}$ the GP yield covariance matrix. This procedure ensures consistency between model and experimental covariance treatments. \\

\bigskip
\textbf{Standardization to experimental space.}
To match the preprocessing applied to the experimental ratios, the model log-ratio means and covariance are standardized using the experimentally determined mean vector $\boldsymbol{\mu}_R^{\mathrm{exp}}$ and standard deviation $\boldsymbol{\sigma}_R^{\mathrm{exp}}$:
\begin{align}
	\boldsymbol{\mathcal{R}}_{\mathrm{std}}^{\mathrm{model}}
	&= \frac{\boldsymbol{\mathcal{R}}^{\mathrm{model}} - \boldsymbol{\mu}_R^{\mathrm{exp}}}
	{\boldsymbol{\sigma}_R^{\mathrm{exp}}}, \\[1mm]
	C_{R}^{\mathrm{std,\,model}}
	&= D^{-1} \, C_R^{\mathrm{model}} \, D^{-1},
	\qquad 
	D = \mathrm{diag}(\boldsymbol{\sigma}_R^{\mathrm{exp}}).
\end{align}
This places the model ratios in the same standardized space used for the PCA of the data.

\bigskip
\textbf{Projection into experimental principal-component space.}
Finally, the standardized model ratios are projected onto the leading $k$ principal components obtained from the experimental PCA. Let $V_k \in \mathbb{R}^{N_R \times k}$ denote the matrix of the top $k$ eigenvectors. The model predictions and uncertainties in PC space are
\begin{align}
	\mathrm{PC}^{\mathrm{model}} 
	&= V_k^\top \, \boldsymbol{\mathcal{R}}_{\mathrm{std}}^{\mathrm{model}}, \\[1mm]
	C_{PC}^{\mathrm{model}}
	&= V_k^\top \, C_{R}^{\mathrm{std,\,model}} \, V_k.
\end{align}
These quantities provide the emulator predictions in the reduced space used for Bayesian calibration.

\subsection{Bayesian Calibration}
\label{subsec:bayesian_calibration}

After validating the GP emulator and mapping all model observables into the experimental principal-component (PC) space, the freeze-out parameters are inferred using Bayesian calibration. 
Compared with conventional $\chi^{2}$ minimization, the Bayesian framework offers several key advantages: it yields the full posterior distribution rather than a single point estimate, incorporates correlated uncertainties in a statistically consistent manner, and mitigates parameter degeneracies that often arise in traditional $\chi^{2}$-based analyses. 
These features make Bayesian inference particularly well suited to the present study, where the observables exhibit strong correlations and the model response is inherently nonlinear. 
Consequently, Bayesian calibration has been widely adopted in heavy-ion physics to address a range of problems, including parameter estimation, model comparison, and uncertainty quantification~\cite{Bernhard:2016tnd,JETSCAPE:2021ehl,Bernhard:2019bmu,Wesolowski:2015fqa,Nijs:2020roc,Parkkila:2021yha}.

The likelihood function quantifies the agreement between the model predictions and the experimental PCs and is modeled as a multivariate Gaussian,
\begin{equation}
	P(\mathcal{D}\mid\theta) \propto 
	\exp\!\left[
	-\tfrac{1}{2}\,
	\big(y_m(\theta)-y_e\big)^{T}
	\Sigma^{-1}
	\big(y_m(\theta)-y_e\big)
	\right],
\end{equation}
where $y_m(\theta)$ denotes the model PCs, $y_e$ the experimental PCs, and $\Sigma$ the total covariance matrix. The covariance $\Sigma$ includes contributions from experimental uncertainties, the GP emulator, and a small regularization term following Ref.~\cite{Bernhard:2018hnz}.

 Flat, non-informative priors are assigned over parameter ranges motivated by previous studies and standard HRG analyses (Table~\ref{tab:params}).
\begin{table}[h!]
	\centering
	\caption{Model parameters and prior ranges used in the Bayesian calibration.}
	\begin{tabular}{ccc}
		\hline
		\textbf{Parameter} & \textbf{Symbol} & \textbf{Range} \\
		\hline
		Temperature & $T$ & $[120,\, 180]~\mathrm{MeV}$ \\
		Baryon chemical potential & $\mu_B$ & $[0,\, 450]~\mathrm{MeV}$ \\
		Strangeness chemical potential & $\mu_S$ & $[-10,\, 150]~\mathrm{MeV}$ \\
		Charge chemical potential & $\mu_Q$ & $[-20,\, 30]~\mathrm{MeV}$ \\
		\hline
	\end{tabular}
	\label{tab:params}
\end{table}

Posterior sampling is performed in PC space using the affine-invariant ensemble Markov Chain Monte Carlo algorithm implemented in \texttt{emcee}~\cite{Foreman-Mackey_2013}. Each run employs 100 walkers initialized uniformly within the prior bounds. A two-stage burn-in of 1,000 steps is followed by 5,000 production steps, yielding roughly $10^{6}$ posterior samples. Convergence is assessed using standard diagnostics: the Gelman--Rubin statistic satisfies $\hat{R}\approx 1.0$ for all parameters, and the integrated autocorrelation times indicate that the effective sample size exceeds 50 times the largest autocorrelation time, providing at least $\mathcal{O}(10^{3})$ effectively independent samples per parameter.
The analysis code used in this work is publicly available at~\cite{HepBayes}

	\section{Results and Discussion}
	\label{sec:results_discussion}
	This analysis uses identified-hadron yield data for, 
	$\pi^{+},\,\pi^{-},\,K^{+},\,K^{-},\,p,\,\bar{p},\,\Lambda,\,\bar{\Lambda},\,\Xi^{-},\,\bar{\Xi}^{+}$ , 
	across collision energies from $\sqrt{s_{NN}} = 7.7$ to $200$~GeV at 
	RHIC~\cite{STAR:2017sal} and $\sqrt{s_{NN}} = 2.76$~TeV at the 
	LHC~\cite{ALICE:2013mez,ALICE:2013cdo,ALICE:2013xmt,Becattini:2014hla}.  
	For each energy, all distinct hadron-yield ratios are constructed following the 
	procedure described in Sec.~\ref{sec:analysis_methodology}.  
	These ratios constitute the inputs to the PCA decorrelation and the Sobol-based 
	global sensitivity analysis presented in the subsections below.

\subsection{Sensitivity Analysis}

\begin{figure*}[t] 
	\centering
	\includegraphics[scale=0.4]{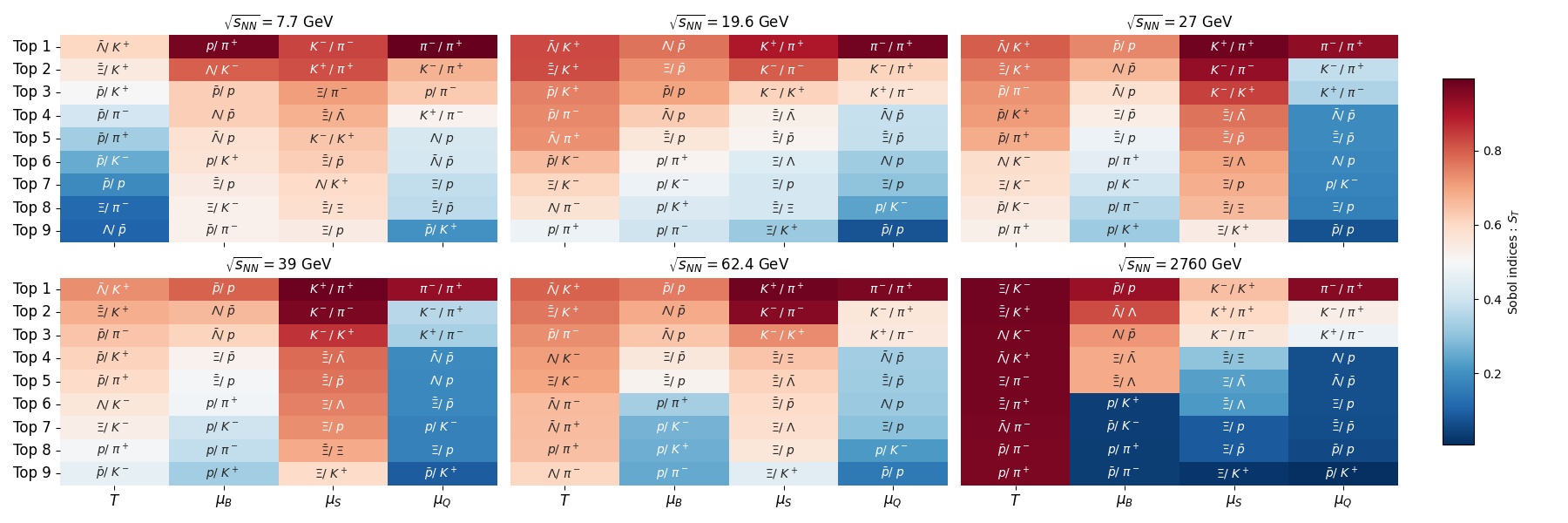}
	\caption{Energy-wise sensitivities of the chemical freeze-out parameters obtained from Sobol total indices ($S_T$). Each column shows the nine algebraically independent hadron yield ratios with the largest $S_T$ values for the corresponding freeze-out parameter. The common color bar indicates the magnitude of $S_T$.}
	\label{fig:sobol_ratio}
\end{figure*}

To quantify how variations in the chemical freeze-out parameters 
$(T, \mu_{B}, \mu_{S}, \mu_{Q})$ influence the predicted hadron-yield ratios, 
we perform a global sensitivity analysis based on Sobol total-effect indices~\cite{Sobol1993, Saltelli2002}.

For a model observable $Y = f(\boldsymbol{\theta})$, the total variance can be 
decomposed as
\[
\mathrm{Var}(Y)=\sum_i V_i + \sum_{i<j} V_{ij} + \cdots + V_{1,2,\dots,d},
\]
where $V_i$ denotes the contribution from parameter $\theta_i$ 
and $V_{ij}$, $V_{ijk}$, etc., denote higher-order interaction terms.
The first-order Sobol index,
\[
S_i = \frac{V_i}{\mathrm{Var}(Y)},
\]
measures the isolated effect of $\theta_i$, while the total index,
\[
S_i^{\mathrm{T}} = 1 - \frac{V_{\sim i}}{\mathrm{Var}(Y)},
\]
captures both the individual and interaction contributions of $\theta_i$.

For each freeze-out parameter and collision energy, we compute the total-effect 
indices $S_i^{\mathrm{T}}$ for all hadron-yield ratios and identify 
the nine algebraically independent ratios with the largest sensitivities.  
These results, summarized in Fig.~\ref{fig:sobol_ratio}, indicate which 
observables carry the dominant information on each thermodynamic parameter and 
how their sensitivity evolves with $\sqrt{s_{NN}}$.

 At $\sqrt{s_{NN}} = 7.7$~GeV, the temperature is the least sensitive parameter, with its influence carried almost entirely by rare strange antibaryon–to–kaon ratios and, to a lesser extent, by antibaryon–to–meson ratios. The small $S_T$ values at this energy reflect the strong suppression of antibaryons—and particularly strange antibaryons—caused by the large $\mu_B$ and $\mu_S$. As $\sqrt{s_{NN}}$ increases, antibaryon suppression weakens, and ratios involving pions, such as $\bar{p}/\pi^{-}$, become increasingly sensitive to $T$. By $\sqrt{s_{NN}} = 27$~GeV, ratios like $\bar{p}/\pi^{-}$ and $\bar{p}/K^{+}$ exhibit $S_T$ values comparable to $\bar{\Lambda}/K^{+}$, indicating the reduced influence of chemical potentials. At the top RHIC energies, many strange and antibaryon ratios act as effective thermal probes. By the LHC regime, $T$ becomes the dominant parameter, and ratios across a wide mass range respond primarily to the Boltzmann factor, making the temperature sensitivity effectively universal.

 In contrast, the Sobol sensitivity of the baryon chemical potential $\mu_B$ decreases monotonically with collision energy, mirroring the gradual reduction in baryon–antibaryon asymmetry. At low energies, $\mu_B$ sensitivity is distributed across both baryon–to–meson ratios (e.g., $p/\pi^+$, $\Lambda/K^-$) and baryon–to–antibaryon ratios (e.g., $\bar{p}/p$, $\Lambda/\bar{p}$). As $\sqrt{s_{NN}}$ increases, the sensitivity becomes concentrated almost entirely in baryon–to–antibaryon ratios such as $\bar{p}/p$, $\bar{\Lambda}/\Lambda$, and $\Xi/\bar{\Lambda}$, while baryon–to–meson ratios lose relevance as their values approach equilibrium limits that are largely insensitive to $\mu_B$. This reflects the near-symmetric production of baryons and antibaryons at high energies, which reduces the influence of $\mu_B$ on the hadronic yields. The strangeness chemical potential $\mu_S$ exhibits dominant sensitivity to kaon–to–pion ratios, with additional contributions at low energies from multi-strange–to–single-strange and non-strange baryon or meson ratios. With increasing energy, a convergence pattern similar to that of $\mu_B$ is observed, although the decline in $\mu_S$ sensitivity occurs more gradually. This slower reduction reflects the progressive approach to global strangeness neutrality in a nearly baryon-symmetric environment, which is effectively realized by LHC energies.

\subsection{Principal Component Analysis}
\label{subsec:pca_energy}

To characterize the dominant structures in the hadron-yield data and their evolution with collision energy, we perform a Principal Component Analysis (PCA) of the log-ratio observables constructed from the experimental yields. To determine which freeze-out parameters control each component, we additionally apply a Sobol global sensitivity analysis to the principal-component amplitudes. The corresponding energy-wise PCA loading matrices and the energy dependence of the principal-component sensitivities are shown in Appendix~\ref{app:pca_details} (Fig.~\ref{fig:PCA_energy_wise}  \ref{fig:PCA_energy_wise} ).

At the lowest RHIC energies (\(\sqrt{s_{NN}} = 7.7\)--19.6~GeV), the leading components are governed primarily by the baryon and strangeness chemical potentials (\(\mu_B,\mu_S\)). These modes capture antibaryon suppression, baryon–meson imbalance, and strangeness–antimatter asymmetry, reflecting the strongly baryon-rich environment. Subleading components encode isospin effects through the \(p/\pi^\pm\) splitting and the temperature dependence of strange-hadron production.

At low intermediate energies (\(27\text{--}39~\mathrm{GeV}\)), the PCA structure shows a clear transition from chemical-potential–dominated to temperature-controlled freeze-out. At \(27~\mathrm{GeV}\), PC1 is still governed by antibaryon–to–meson scaling set by \(\mu_B\) and \(\mu_S\), reflecting the reduced net-baryon density and enhanced antibaryon production. PCs~2--4, dominated by pion-normalized baryon and kaon ratios, exhibit mixed sensitivity to \(T\) and \(\mu_S\), signaling the onset of thermal mass ordering, while the higher modes further separate chemical and thermal effects, most prominently in strange-baryon–to–kaon ratios such as \(\Xi/K^+\) and \(\Lambda/K^+\). By \(39~\mathrm{GeV}\), PCs~1--2 become largely temperature driven, governed by \(\Xi\) and \(\bar{\Xi}\) yields, whereas the remaining components retain only minor charge and strangeness structure. The diminishing role of \(\mu_B\) and \(\mu_S\) thus marks the emergence of a nearly thermally equilibrated freeze-out surface.

At the high RHIC energies (\(62.4~\mathrm{GeV}\)) and above, temperature largely controls the hadron-yield pattern. The leading PCs are dominated by multistrange and antibaryon ratios (\(\Xi\), \(\bar{\Lambda}\), \(\bar{p}\)), reflecting the near restoration of matter–antimatter symmetry at these energies. Intermediate components capture residual charge–strangeness correlations and baryon–meson balance, while higher modes contain only small fluctuations from weak chemical-potential effects. At the LHC energy of \(\sqrt{s_{NN}} = 2.76~\mathrm{TeV}\), freeze-out becomes almost purely thermal: PCs~1--2 are governed by \(\Lambda\)- and \(\bar{\Lambda}\)-related ratios, with the remaining components contributing negligibly. This corresponds to a regime of effectively vanishing chemical potentials, where hadron yields are set primarily by the temperature.

Overall, the PCA decorrelates the ratio space into nine statistically independent directions whose energy dependence reveals a smooth transition from a baryon-rich, chemically driven regime to a nearly thermal, temperature-dominated freeze-out.

It is worth noting that interpreting individual principal components can be challenging because each PC mixes several correlated observables. To aid interpretation, we identify at each collision energy the nine algebraically independent hadron-yield ratios with the largest absolute loadings across all PCs; the corresponding figures are provided in Appendix~\ref{app:pca_details} (Fig.~\ref{fig:loading_pca_energy_wise}). These ratios span directions in observable space that carry maximal variance, making them the most informative and physically constraining combinations in the dataset. Although algebraically independent, they are not necessarily physically independent. For example, at \(\sqrt{s_{NN}}=2.76~\mathrm{TeV}\), the ratios \(\Lambda/\bar{p}\) and \(\bar{\Lambda}/\bar{p}\) remain correlated through antibaryon production and baryon–strangeness conservation, while ratios such as \(\Xi/K^+\) and \(\Lambda/K^-\) share common strange-baryon and kaon production mechanisms. These examples underscore the need for PCA to isolate the least redundant and most informative observable combinations, even when physical correlations persist.

\subsection{MCMC Calibration}
\label{subsec:mcmc_calibration} 

The HRG model parameters were calibrated independently at each collision energy from $\sqrt{s_{NN}} = 7.7$ to $200~\mathrm{GeV}$ at RHIC~\cite{STAR:2017sal} and at $\sqrt{s_{NN}} = 2.76~\mathrm{TeV}$ at the LHC~\cite{ALICE:2013mez,ALICE:2013cdo,ALICE:2013xmt,Becattini:2014hla} using Bayesian inference. All analyses were performed in the principal-component (PC) space constructed from the logarithmic hadron-yield ratios in the most central collisions. As discussed in Sec.~\ref{subsec:pca_energy}, the first nine PCs contain all independent information from the ratio ensemble and therefore form the basis for the parameter calibration.

Following earlier studies~\cite{Bhattacharyya:2019cer,Bhattacharyya:2019wag,Bhattacharyya:2020sgn,STAR:2017sal}, two complementary fitting schemes were employed to examine the sensitivity of the extracted parameters to the imposed conservation constraints.

\textbf{Case 1: Unconstrained fit.}  
All chemical freeze-out parameters---the temperature ($T$), baryon chemical potential ($\mu_B$), strangeness chemical potential ($\mu_S$), and charge chemical potential ($\mu_Q$)---were treated as free parameters in the Bayesian calibration.  
This provides the most general determination of the freeze-out conditions without explicitly enforcing any external constraints.
 
\textbf{Case 2: Constrained fit.}  
The strangeness and electric-charge chemical potentials were fixed by imposing global strangeness neutrality and a constant charge-to-baryon ratio~\cite{Alba:2014eba}:
These conditions,
\begin{align}
	\sum_i n_i(T, \mu_B, \mu_S, \mu_Q)\, S_i &= 0, \label{eq:strangeness_constraint} \\
	\sum_i n_i(T, \mu_B, \mu_S, \mu_Q)\, Q_i &= r\,\sum_i n_i(T, \mu_B, \mu_S, \mu_Q)\, B_i, \label{eq:charge_constraint}
\end{align}
These constraints reduce the fit to two independent parameters, $T$ and $\mu_B$, with $\mu_S$ and $\mu_Q$ determined implicitly from Eqs.~\eqref{eq:strangeness_constraint}–\eqref{eq:charge_constraint}.

\begin{figure*}[t]
	\centering
	\includegraphics[scale=0.4]{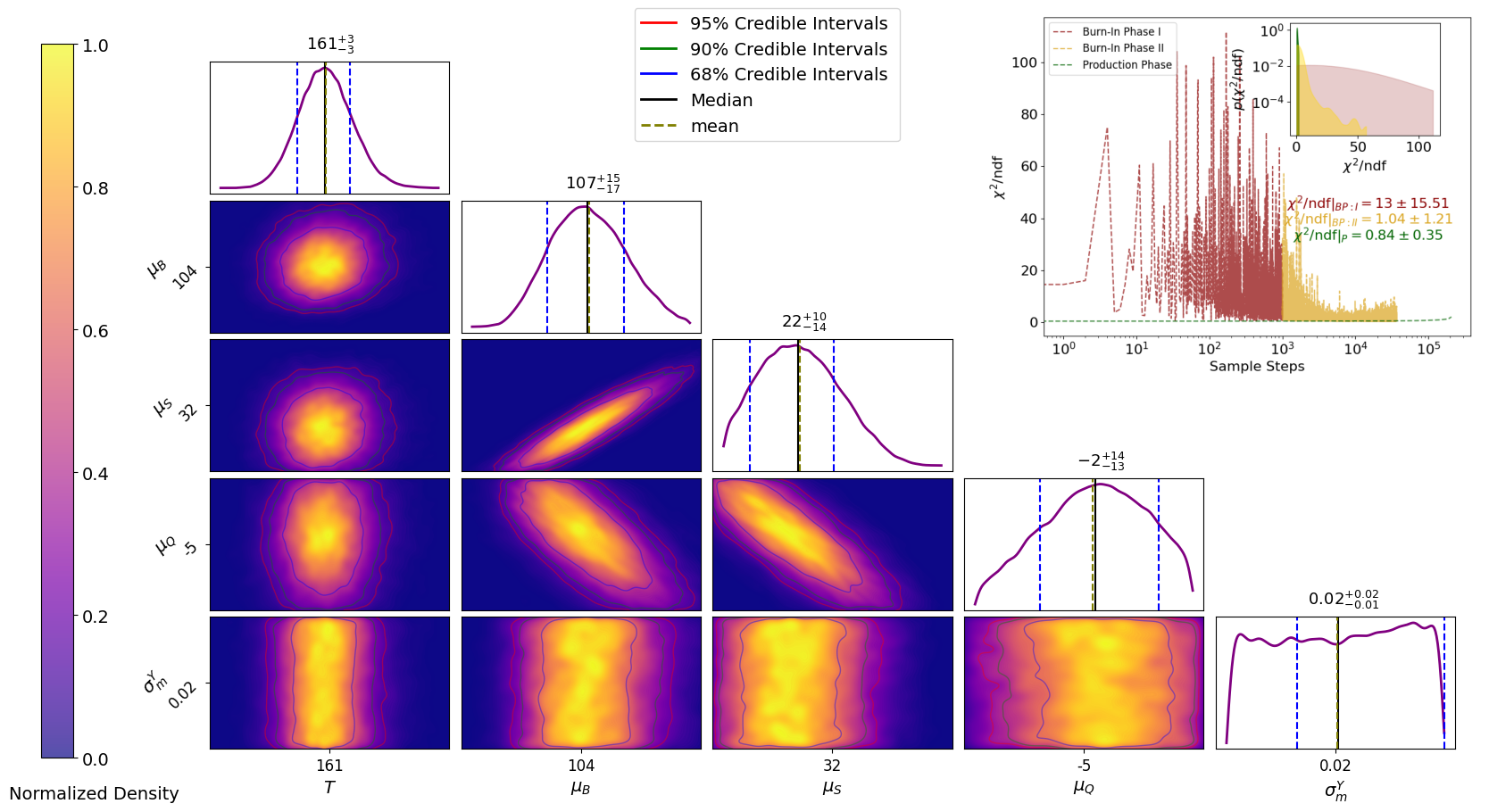}
	\caption{
		Posterior distributions from the Bayesian calibration at $\sqrt{s_{NN}}=39$~GeV (Case~I).
		Diagonal: one-dimensional marginalized posteriors with mean (dashed) and median (solid) lines.
		Off-diagonal: two-dimensional joint posteriors with 68
		Inset: evolution of $\chi^2/\mathrm{ndf}$ during burn-in and production sphase.
	}
	\label{fig:cornar_plot}
\end{figure*}

  \begin{figure*} 
 	\centering
 	\includegraphics[scale=0.35]{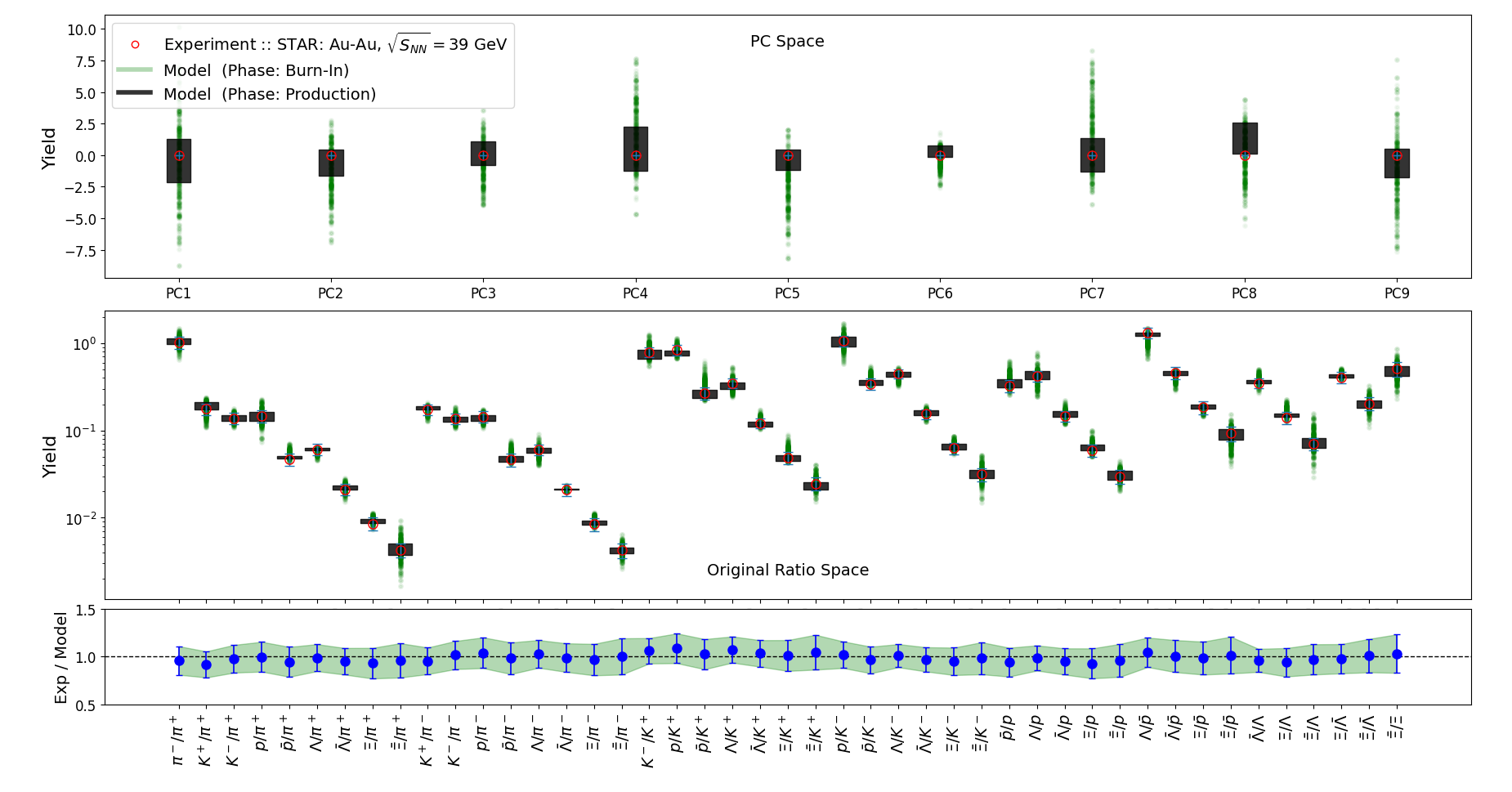}

 \caption{
 	Comparison between experimental data and the Bayesian-calibrated HRG model for $\sqrt{s_{NN}} = 39$~GeV (Case~I).  
 	\textbf{Top:} Experimental data (red circles) and calibrated model predictions (black bands) for the nine dominant principal components (PCs) of hadron-yield ratios.  
 	The green bands denote samples from the MCMC burn-in phases.  
 	\textbf{Middle:} Comparison between the calibrated model predictions and the original experimental hadron-yield ratios, obtained by transforming the model results from the decorrelated PC space back into the physical ratio space using the inverse PCA transformation.  
 	\textbf{Bottom:} Ratio of experimental to marginalized model predictions for all 45 hadron-yield ratios, demonstrating excellent agreement across the full observable set.  
 }

 	\label{fig:exp_model_comp}
 \end{figure*}

 The detailed results of the Bayesian calibration for both approaches (Case~I and Case~II) are summarized in Table~\ref{tab:par_value}. The MCMC posterior distributions for Case~I at $\sqrt{s_{NN}} = 39$~GeV are shown in Fig.~\ref{fig:cornar_plot}, which displays the one-dimensional marginalized posteriors along the diagonal and the two-dimensional parameter correlations in the off-diagonal panels. The posterior structure shows that the freeze-out temperature $T$ is largely uncorrelated with the chemical potentials. In the HRG model, $T$ affects all hadron species in a broadly similar way, whereas $\mu_B$, $\mu_S$, and $\mu_Q$ primarily control the relative baryon, antibaryon, and strangeness abundances. As a result, variations in the chemical potentials have limited impact on the temperature preferred by the data, yielding a comparatively narrow posterior for $T$.
 
 In contrast, the chemical potentials exhibit strong mutual correlations that arise naturally from the composition of the hadronic spectrum. An increase in $\mu_B$ enhances baryon yields relative to antibaryons, which in turn requires a larger $\mu_S$ to maintain the observed ratios of strange baryons to mesons, producing a positive $\mu_B$–$\mu_S$ correlation. Conversely, $\mu_Q$ is strongly anti-correlated with both $\mu_B$ and $\mu_S$: increases in baryon number or strangeness content raise the abundance of positively charged hadrons, necessitating a compensating reduction in $\mu_Q$ to maintain the observed charge balance. It is worth noting that these interdependencies are not imposed externally but emerge from the statistical interplay between baryon number, strangeness, and electric charge conservation within the HRG framework. The resulting posterior correlations provide clear physical insight into parameter sensitivities and degeneracies in the freeze-out description. They also explain why enforcing explicit strangeness neutrality and a fixed charge-to-baryon ratio in Case~II leads to only minor shifts relative to the unconstrained Case~I: the experimental data already encode these conservation relations through the correlated posterior structure.

 The top-right inset of Fig.~\ref{fig:cornar_plot} shows the evolution of $\chi^{2}/\mathrm{ndf}$ during the MCMC sampling, where 
 $\chi^{2} = (y_{\mathrm{exp}} - y_{\mathrm{model}})^{\top}\Sigma^{-1}(y_{\mathrm{exp}} - y_{\mathrm{model}})$, and $\Sigma$ denotes the total covariance matrix, incorporating uncertainties from both the model predictions and the experimental measurements. A rapid decrease in $\chi^2/\mathrm{ndf}$ during the second burn-in period, followed by stable fluctuations around a constant mean during production,further confirms the convergence and equilibration of the Markov chains. The steady mean value, $\chi^2/\mathrm{ndf} = 0.75 \pm 0.51$, indicates that the HRG model provides a statistically consistent description of the data within the posterior credible intervals. It should be noted, however, that this $\chi^2/\mathrm{ndf}$ is evaluated in the principal-component (PC) space, where the observables are orthogonalized and decorrelated, rather than from a direct comparison of hadron-yield ratios.  Hence, it quantifies the model’s agreement with the data in the reduced, information-preserving PC basis. For completeness, the corresponding $\chi^2/\mathrm{ndf}$ values obtained from direct comparisons of experimental and model-predicted hadron ratios are summarized in Table~\ref{tab:par_value}, providing a complementary validation in the physical observable space.

\begin{figure*}[t]
	\centering
	\includegraphics[scale=0.4]{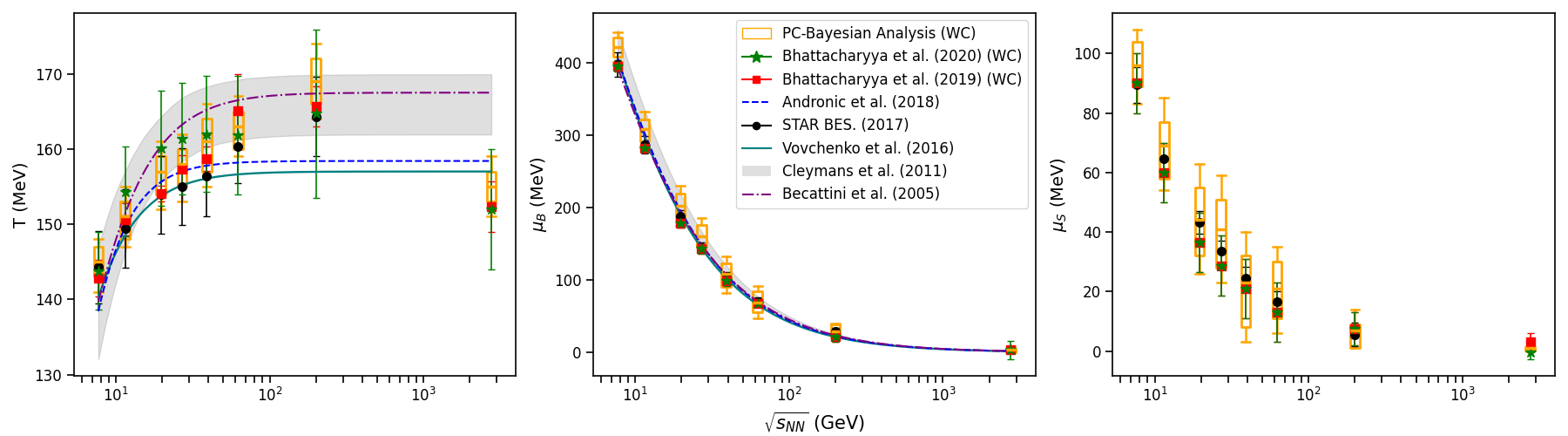}
	\caption{
		Comparison of the chemical freeze-out parameters extracted without constraints (WC) in the present PCA-based Bayesian analysis (orange box--whisker symbols) with earlier determinations 
		\cite{Bhattacharyya:2019cer,Bhattacharyya:2019wag,Andronic:2017pug,STAR:2017sal,Vovchenko:2015idt,Cleymans:2005xv,Becattini:2005xt} across a range of collision energies. 
		Each orange box denotes the 68\% credible interval, the horizontal line marks the posterior median, and the whiskers show the 95\% credible interval.  
		\textbf{Left:} Freeze-out temperature ($T$).  
		\textbf{Middle:} Baryon chemical potential ($\mu_B$).  
		\textbf{Right:} Strangeness chemical potential ($\mu_S$).
	}
	\label{fig:freezeout_comparison}
\end{figure*}

The comparison between the calibrated HRG model and the experimental data for $\sqrt{s_{NN}}=39$~GeV in \textbf{Case~I} is shown in Fig.~\ref{fig:exp_model_comp}; the figure caption provides additional details. The top panel compares the nine dominant principal components (PCs) of the experimental points with the calibrated model. The close overlap between the data and the posterior bands shows that the model accurately reproduces the variance structure in PC space. The middle panel compares the calibrated model predictions with all 45 constructed experimental hadron-yield ratios in the original ratio space and shows excellent agreement. As confirmed in the bottom panel, the data-to-model ratios cluster tightly around unity across all observables. It is worth emphasizing that these ratios were not used directly in the fit: the calibration was performed entirely in the principal-component (PC) basis, and the values shown here are obtained by transforming the model predictions back to ratio space using the inverse PCA map. This demonstrates that the PCA-based Bayesian framework retains the full information content of the observables while removing redundancy from correlated ratios, achieving statistical consistency comparable to a full multivariate fit in ratio space but with improved numerical stability and transparent uncertainty propagation.

The quantitative results of the Bayesian calibration, summarized in Table~\ref{tab:par_value}, show that the freeze-out parameters obtained from both approaches agree closely across all collision energies. This consistency indicates that imposing global strangeness neutrality and a fixed charge-to-baryon ratio provides a physically motivated reduction of the parameter space without degrading the fit quality. Although the central values of $\mu_S$ and $\mu_Q$ are nearly identical in the two cases, their statistical interpretation differs: in Case~I they are treated as free parameters sampled directly in the MCMC analysis, whereas in Case~II they are determined implicitly from the neutrality conditions, with their variations arising solely from fluctuations in $T$ and $\mu_B$. The slightly narrower posterior widths in Case~II reflect the constraining effect of these physical relations, which reduce parameter degeneracies and improve the precision of the extracted freeze-out parameters.

A variety of methodologies have been employed to extract chemical freeze-out parameters within the HRG framework, differing in fitting strategies, model assumptions, and the choice of hadron ratios. Several studies introduce a strangeness–saturation parameter $\gamma_s$ to account for possible deviations from full strangeness equilibration. For example, Refs.~\cite{Andronic:2005yp,STAR:2017sal} analyze a restricted subset of particle ratios while simultaneously fitting $\gamma_s$. In contrast, Ref.~\cite{Bhattacharyya:2019cer} examines the sensitivity of the extracted parameters to ratio selection by performing independent fits with nine representative ratio sets, whereas Ref.~\cite{Bhattacharyya:2020sgn} undertakes an exhaustive exploration of all possible hadron-ratio combinations (of order $10^8$) to quantify the systematic uncertainties associated with ratio choice.

A comparison of the extracted freeze-out parameters from this work with previous analyses across collision energies is presented in Fig.~\ref{fig:freezeout_comparison}.  
The results obtained using both fitting strategies (Case~I and Case~II) show excellent overall agreement with earlier determinations based on diverse methodological approaches.  
The mean values of the parameters extracted in this study are slightly higher than those reported in most earlier works \cite{Andronic:2005yp,STAR:2017sal,Bhattacharyya:2019cer,Vovchenko:2015idt} but remain well within the corresponding credible intervals. In particular, at $\sqrt{s_{NN}} = 200~\mathrm{GeV}$, our results show remarkable consistency with the comprehensive analysis of Ref.~\cite{Bhattacharyya:2020sgn}, further validating the reliability of the present Bayesian–PCA calibration framework.  
Consistent with previous HRG-based studies, the freeze-out temperature $T_{\mathrm{CF}}$ is more tightly constrained than the chemical potentials, reflecting its dominant role in determining the overall hadron yield distribution.

The posterior distributions show that $\mu_S$ and $\mu_Q$—particularly $\mu_Q$—exhibit broader credible intervals across collision energies than $\mu_B$ and $T$. Among the chemical potentials, $\mu_Q$ carries the largest relative uncertainty, reflecting its strong compensating correlations with $\mu_B$ and $\mu_S$. This trend is consistent with the very small Sobol sensitivity indices for $\mu_Q$ (Fig.~\ref{fig:Sobol_PCA_plot}), indicating its weak direct influence on the modeled observables. In comparison, Ref.~\cite{Bhattacharyya:2019wag} reports narrower uncertainties for $\mu_S$ and $\mu_Q$, though the central values remain consistent with our results. That study determines the freeze-out parameters from a reduced set of conserved-charge observables rather than fitting the full set of hadron-yield ratios. The correspondingly smaller error bands likely arise from direct $\chi^{2}$ minimization with fixed observables and a more limited treatment of uncertainties. In contrast, the Bayesian–PCA framework developed here propagates experimental correlations, emulator variance, and statistical fluctuations, yielding broader but more reliable credible intervals that better reflect the true information content of the data.

\begin{table*}[h!]
	\centering
	\caption{
		Posterior mean, median, and credible intervals (CI) for the calibrated freeze-out parameters at various collision energies $\sqrt{s_{NN}}$. The left and right tables correspond to \textbf{Case~I} and \textbf{Case~II}, respectively. In \textbf{Case~II}, $\mu_S$ and $\mu_Q$ are not fitted parameters but are determined from the fitted $T$ and $\mu_B$ using Eqs.~(\ref{eq:strangeness_constraint})–(\ref{eq:charge_constraint}). The quantity $\chi^{2}/\mathrm{ndf}$ is computed by comparing the model predictions with the experimental values in the experimental ratio space. All quantities are in MeV.
	}
	\begin{minipage}{0.48\linewidth}
		\label{tab:par_value}
		\centering
		\resizebox{\linewidth}{!}{
			\begin{tabular}{lcccccc}
				\toprule
				Parameter & Mean & Median & MAP& 68\% CI & 90\% CI & 95\% CI   \\
				\midrule
				\multicolumn{7}{c}{\textbf{7.7 GeV \ \ ($\chi^2$/ndf = 2.12 $\pm$ 0.77)} } \\
				\midrule
				$T$     & 145 & 144 &143 & 143--147 & 141--148 & 141--149  \\
				$\mu_{B}$ & 422 & 421 &394 &409--435 & 402--443 & 399--447   \\
				$\mu_{S}$ & 96  & 96  &74& 89--104   & 83--108   & 81--111    \\
				$\mu_{Q}$ & -16   & -15  &-12 & $>$-17   & $>$-23  & $>$-25     \\
				$\sigma^y_m$ & 0.015 & 0.014 &0.009& 0.004--0.020 & 0.002--0.029 & 0.001--0.034   \\
				\midrule
				
				\multicolumn{7}{c}{\textbf{11.5 GeV \ \ ($\chi^2$/ndf = 1.91 $\pm$ 0.51)} } \\
				\midrule
				$T$     & 151 & 151 & 150& 148--153 & 147--155 & 146--156   \\
				$\mu_{B}$ & 307 & 306 &292 & 290--321 & 281--331 & 277--336   \\
				$\mu_{S}$ & 68  & 66  & 56 & 56--75   & 50--82   & 49--86     \\
				$\mu_{Q}$ & -6  & -5  & -5& -8 -- 4   & $>$-17    & $>$-22     \\
				$\sigma^y_m$         &  0.005 & 0.001 & 0.001& 0.001--0.002 & $<$0.003 & $<$0.005  \\
				\midrule
				\multicolumn{7}{c}{\textbf{19.6 GeV \ \ ($\chi^2$/ndf = 1.87 $\pm$ 0.60)} } \\
				\midrule
				$T$     & 157 & 157 & 156&154--159 & 152--161 & 152--162   \\
				$\mu_{B}$ & 203 & 202 &191 & 185--219 & 175--230 & 172--237   \\
				$\mu_{S}$ & 44  & 43  &35 & 32--55   & 26--63   & 23--67    \\
				$\mu_{Q}$ & -5  & -3  &-4 & $>$-9   & $>$ -20   & $>$-24     \\
				$\sigma^y_m$           & 0.020 &  0.019 &0.033 &0.003--0.029 & 0.002--0.037 & $<$0.038  \\
				\midrule
				\multicolumn{7}{c}{\textbf{27 GeV \ \ ($\chi^2$/ndf = 1.69 $\pm$ 0.76)} } \\
				\midrule
				$T$     & 158 & 158 &158 &156--161 & 154--162 & 153--163  \\
				$\mu_{B}$ & 154 & 153 &148 &136--170 & 126--181 & 122--186 \\
				$\mu_{S}$ & 34  & 33  &30 &20--46   & 13--54   & 10--60     \\
				$\mu_{Q}$ & -2  & -2  &-2 & $>$-6  & $>$-17   & $>$-22    \\
				$\sigma^y_m$          & 0.020 & 0.020 & 0.03&0.012--0.030 & $<$0.035 & $<$0.038   \\
				\midrule

				\multicolumn{7}{c}{\textbf{39 GeV \ \ ($\chi^2$/ndf = 0.75 $\pm$ 0.51)} } \\
				\midrule
				$T$       & 161 & 161 &160 & 157--164 & 155--166 & 154--167  \\
				$\mu_{B}$ & 108 & 107 &102 & 91--123  & 82--133  & 79--139     \\
				$\mu_{S}$ & 23  & 22  &18 & 8--32    & 3--40    & 1--43      \\
				$\mu_{Q}$ & -3  & -2  &-2 & -15--12  & -21--19  & -24--20    \\
				$\sigma^y_m$          & 0.020 & 0.021  &0.030 & 0.013--0.040 & 0.004--0.042 & 0.002--0.042   \\
				\midrule
				
				\multicolumn{7}{c}{\textbf{62.4 GeV \ \ ($\chi^2$/ndf = 1.52 $\pm$ 0.65)} } \\
				\midrule
				$T$     & 163 & 163 &162 &160--165 & 159--167 & 158--168   \\
				$\mu_{B}$ & 69  & 69  &66 & 55--84   & 47--92   & 44--96    \\
				$\mu_{S}$ & 21  & 20  &18 & 11--30   & 6--35    & 3--37     \\
				$\mu_{Q}$ & -3  & -3  &-1 & -9--8    & $>$-15   & $>$-17    \\
				$\sigma^y_m$          & 0.020 & 0.019 & 0.032& 0.003--0.032 & 0.002--0.038 & 0.002--0.040   \\
				\midrule
				
				\multicolumn{7}{c}{\textbf{200 GeV \ \ ($\chi^2$/ndf = 0.52 $\pm$ 0.26)} } \\
				\midrule
				$T$     & 169 & 169 &169 &166--172 & 164--174 & 164--175  \\
				$\mu_{B}$ & 29  & 30  &26 & 25--39   & 20--40   & 17--40     \\
				$\mu_{S}$ & 7   & 7   &4 & 1--9     & $<$14    & $<$15       \\
				$\mu_{Q}$ & -1  & -1  &-1 & -10--2   & $<$7   & $<$8      \\
				$\sigma^y_m$          & 0.021 & 0.021 &0.038& 0.013--0.039 & 0.004--0.040 & 0.002--0.040   \\
				\midrule
				
				\multicolumn{7}{c}{\textbf{2760 GeV \ \ ($\chi^2$/ndf = 0.81 $\pm$ 0.38)} } \\
				\midrule
				$T$     & 155 & 155 &155 &152--157 & 151--159 & 150--160  \\
				$\mu_{B}$ & 4  & 4  &0.1 & 0.01--5   & 0--8   & 0--9     \\
				$\mu_{S}$ & 0.40   & 0.46   & 0.01 & -2.0 -- 3.8     & -3.3 -- 4.0    &  -5.0-- 5.0       \\
				$\mu_{Q}$ & -0.43  & 0.55  & 0.0 & -4.0 --1.0   & -5.0--3.5   & $<4.2$     \\
				$\sigma^y_m$          & 0.020 & 0.021 &0.005& 0.013--0.030 & $<$0.035 & $<$0.040   \\
				\midrule
				
			\end{tabular}
			
		}

	\end{minipage}
	\hfill
	\begin{minipage}{0.48\linewidth}
		\centering
		\resizebox{\linewidth}{!}{
			\begin{tabular}{lcccccc}
				\toprule
				Parameter & Mean & Median  & MAP & 68\% CI & 90\% CI & 95\% CI   \\
				\midrule
				\multicolumn{7}{c}{\textbf{7.7 GeV \ \ ($\chi^2$/ndf = 1.97 $\pm$ 0.74)} } \\
				\midrule
				$T$     & 143 & 143 &143& 141--145 & 140--146 & 139--147   \\
				$\mu_{B}$ & 419 & 419 &420& 407--431 & 398--439 & 396--444   \\
				$\mu_{S}$ & 100  & 100  & -- & 93--106   & 92--113   & 88--113    \\
				$\mu_{Q}$ & -11  & -11  & -- & -12 -- -10   & -12 -- -10   & -12 -- -10    \\
				$\sigma^y_m$ & 0.015 & 0.013 &0.010& 0.045--0.019 & 0.002--0.029 & 0.001--0.033 \\
				\midrule
				
				\multicolumn{7}{c}{\textbf{11.5 GeV \ \ ($\chi^2$/ndf = 1.68 $\pm$ 0.31)} } \\
				\midrule
				$T$     & 150 & 150 & 148--152 & 147--154 & 146--155 & 145--156 \\
				$\mu_{B}$ & 311 & 310 & 297--323 & 288--331 & 286--337 & 279--345 \\
				$\mu_{S}$ & 71  & 70  & 65--76   & 61--80   & 60--83   & 58--89 \\
				$\mu_{Q}$ & -8  & -8  & -9-- -7   & -9-- -7  & -10---7  & -10-- -7 \\
				$\sigma^y_m$          & 0.014 & 0.001 &0.001& 0.001--0.002 & 0.001--0.004 &    $<$0.010   \\
				\midrule
				\multicolumn{7}{c}{\textbf{19.6 GeV \ \ ($\chi^2$/ndf = 1.60 $\pm$ 0.51)} } \\
				\midrule
				$T$     & 156 & 156 &156 & 154--158 & 152--160 & 151--161   \\
				$\mu_{B}$ & 204 & 204 &203 & 194--213 & 189--220 & 186--223  \\
				$\mu_{S}$ & 47  & 47  &-- & 43--50   & 40--53   & 39--54   \\
				$\mu_{Q}$ & -6  & -6  &-- & -6-- -5   & -6-- -5  & -6-- -5   \\
				$\sigma^y_m$   & 0.020 & 0.020 &0.003 & $<$0.028 & $<$0.036 & $<$  0.038   \\
				\midrule

				\multicolumn{7}{c}{\textbf{27 GeV \ \ ($\chi^2$/ndf = 1.33 $\pm$ 0.48)} } \\
				\midrule
				$T$     & 159 & 159 &158 &156--161 & 154--163 & 154--164  \\
				$\mu_{B}$ & 155 & 155 &155 & 146--163 & 141--169 & 139--172   \\
				$\mu_{S}$ & 35  & 35  & -- & 32--39   & 29--42   & 27--44      \\
				$\mu_{Q}$ & -4  & -4  & -- & -5-- -4  & -5-- -3  & -5-- -3    \\
				$\sigma^y_m$          & 0.020 & 0.021 & 0.002& 0.013--0.039 & $<$0.04 & $<$0.055   \\
				\midrule
				\multicolumn{7}{c}{\textbf{39 GeV \ \ ($\chi^2$/ndf = 0.33 $\pm$ 0.23)} } \\
				\midrule
				$T$       &160   & 160 &160 & 157--163 & 155--166 & 154--167   \\
				$\mu_{B}$ & 111 & 111 &110 &  100--121  & 94--128  & 91--132     \\
				$\mu_{S}$ & 26  & 26  & --&  22--28    & 20--30    & 20--32         \\
				$\mu_{Q}$ & -3  & -3  & -- &  -3-- -3  & -4-- -2  & -4-- -2     \\
				$\sigma^y_m$          & 0.03& 0.020 & 0.021 & 0.001--0.027 & 0.003--0.036 & 0.001--0.038   \\
				\midrule
				\multicolumn{7}{c}{\textbf{62.4 GeV \ \ ($\chi^2$/ndf = 1.10 $\pm$ 0.28)} } \\
				\midrule
				$T$     & 162 & 162 &162 & 159--164 & 158--165 & 158--166   \\
				$\mu_{B}$ & 60  & 60  &61 & 52--69   & 46--74   & 43--77    \\
				$\mu_{S}$ & 14  & 14  &-- & 13--16   & 11--17    & 10--17        \\
				$\mu_{Q}$ & -2  & -2  &-- & -2--2    & -2-- -1  & -2-- -1     \\
				$\sigma^y_m$          & 0.020 & 0.021 &0.007 & $<$0.027 & $<$0.036 &  $<$0.038   \\
				\midrule
				
				\multicolumn{7}{c}{\textbf{2760 GeV \ \ ($\chi^2$/ndf = 0.42 $\pm$ 0.17)} } \\
				\midrule
				$T$     & 169 & 169 & 169 & 166--172 & 164--174 & 164--175   \\
				$\mu_{B}$ & 31  & 31  & 31 & 22--40   & 16--45   & 14--48     \\
				$\mu_{S}$ & 6   & 7   & -- & 5--9     & 2--10    & 1--10       \\
				$\mu_{Q}$ & -1  & -1  & -- & -1.2 -- -0.5   & -1.2-- -0.2   & -1.3--0.0     \\
				$\sigma^y_m$         & 0.022 & 0.023 &0.007& 0.013--0.028 & $<$0.039 & $<$ 0.041  \\
				\midrule
				
				\multicolumn{7}{c}{\textbf{2760 GeV \ \ ($\chi^2$/ndf = 0.55 $\pm$ 0.24)} } \\
				\midrule
				$T$     & 155 & 155 & 155 & 153--157 & 151--159 & 150--160   \\
				$\mu_{B}$ & 4  & 4  & 1 & 0.01-- 5.0  & 0.01 -- 8.0   & 0.0 --9.0    \\
				$\mu_{S}$ & 1   & 0.9   & -- & 0.20 -- 1.3   &  0.04 --1.71   & 0.0 -- 1.8    \\
				$\mu_{Q}$ & -0.1  & -0.1  & -- & -0.16 -- -0.02  & -0.20 -- -0.01  & -0.22 -- 0.0     \\
				$\sigma^y_m$ &0.020& 0.022 & 0.002 & 0.01 -- 0.03 & 0.003-- 0.04 & 0.002--0.04    \\
				\midrule

			\end{tabular}
		}

	\end{minipage}
\end{table*}

\section{Summary and Conclusions}
\label{sec:conclusion}

This work presents a statistically consistent, data-driven framework for determining the chemical freeze-out parameters in ultra-relativistic heavy-ion collisions using a Principal Component Analysis (PCA)–based Bayesian calibration of the Hadron Resonance Gas (HRG) model. The approach addresses limitations of conventional ratio-based analyses by eliminating redundancy among hadron ratios and ambiguities arising from arbitrary ratio selection with proper treatment of correlations.

Starting from the identified hadron yields at each collision energy—from $\sqrt{s_{NN}}=7.7$~GeV at RHIC to $2.76$~TeV at the LHC—all possible yield ratios were constructed and transformed into an orthogonal basis using PCA. The Bayesian calibration was performed directly in this decorrelated principal-component space—rather than the raw, correlated ratio space, ensuring that only the independent degrees of freedom enter the analysis. A Gaussian Process (GP) emulator, validated against full HRG calculations, provided efficient and accurate predictions across the four-dimensional parameter space $(T,\mu_B,\mu_S,\mu_Q)$ throughout the MCMC sampling.

An energy-dependent global Sobol sensitivity analysis was performed to identify the nine algebraically independent hadron-yield ratios most sensitive to the freeze-out parameters. The results show a clear transition from chemical-potential–dominated to temperature-controlled freeze-out with increasing $\sqrt{s_{NN}}$. At low energies, the dominant sensitivities are to $\mu_B$ and $\mu_S$, reflecting strong baryon–antibaryon asymmetry and strangeness suppression; $\mu_B$ sensitivity appears in both baryon–to–meson and baryon–to–antibaryon ratios. With increasing energy, these effects weaken and temperature becomes the primary control parameter, with ratios such as $\bar{\Lambda}/K^+$ and $\bar{\Xi}/K^+$ acting as effective thermal probes. By LHC energies, the system is nearly matter–antimatter symmetric and strangeness neutral, so the chemical potentials contribute only marginally. Consistently, with increasing energy, the $\mu_B$ sensitivity becomes restricted to baryon–to–antibaryon ratios, while the $\mu_S$ sensitivity—dominated by kaon–to–pion ratios—decreases more gradually as the system approaches global strangeness neutrality.

We have also performed a rigorous energy-dependent Principal Component Analysis (PCA) of experimental hadron-yield ratios, complemented by a Sobol sensitivity analysis of the principal components.  The resulting energy-wise PCA loading maps and parameter sensitivities were presented to identify the dominant observables and to trace the transition from chemical-potential–dominated to temperature-controlled freeze-out dynamics.

The HRG parameters were calibrated at each energy using the Bayesian–PCA framework, which performs the analysis entirely in decorrelated PC space to retain all experimental information without redundancy. Although individual ratios are not fitted directly, the reconstructed predictions agree well with the data, with data-to-model ratios clustering around unity for all 45 observables. The posteriors show that $T$ is largely uncorrelated with the chemical potentials, while $\mu_B$, $\mu_S$, and $\mu_Q$ exhibit strong mutual correlations arising from baryon number, charge, and strangeness conservation. The broader credible intervals for $\mu_S$ and especially $\mu_Q$ reflect their weaker direct influence on the observables, consistent with their small Sobol sensitivities. The results from the unconstrained and constrained fits are in close agreement across all energies, indicating that the conservation conditions are already encoded in the data. The extracted freeze-out parameters are consistent with previous HRG-based studies, confirming the robustness of the Bayesian–PCA framework.

In conclusions, the PCA–Bayesian framework introduced here provides a statistically consistent method for determining chemical freeze-out conditions. By working in a PCA-decorrelated space, the approach eliminates systematic uncertainties associated with ratio selection and ensures that only the independent information in the data enters the calibration. The Bayesian formulation enables a transparent and uncertainty-aware extraction of the freeze-out parameters, while the combined PCA loadings and Sobol sensitivity analysis offer direct physical insight into which hadron ratios, and which ratio combinations, are most informative across collision energies. Looking ahead, this framework can be extended to centrality-dependent analyses that incorporate correlations between centrality bins, as well as to studies of sequential freeze-out, higher-order fluctuations, and interacting HRG formulations, including QCD-motivated extensions of the hadron spectrum.
 \section{ACKNOWLEDGEMENTS}

 The author gratefully acknowledges Deeptak Biswas and Rajesh Biswas for their careful reading of the manuscript and for many valuable suggestions. Support from the Department of Physics, Amrita Vishwa Vidyapeetham, is sincerely appreciated. The author also wishes to express deep gratitude to Tanu Day for her unwavering and unconditional support throughout this work.

	\appendix
\section{Extended Energy-Dependent PCA and Sobol Sensitivity Analysis}

	\label{app:pca_details}
	 
	This appendix collects the extended figures associated with the energy-dependent PCA and the Sobol global sensitivity analysis discussed in Sec.~\ref{subsec:pca_energy}.
	
\begin{figure*}[t]
		\centering
		\includegraphics[scale=0.4]{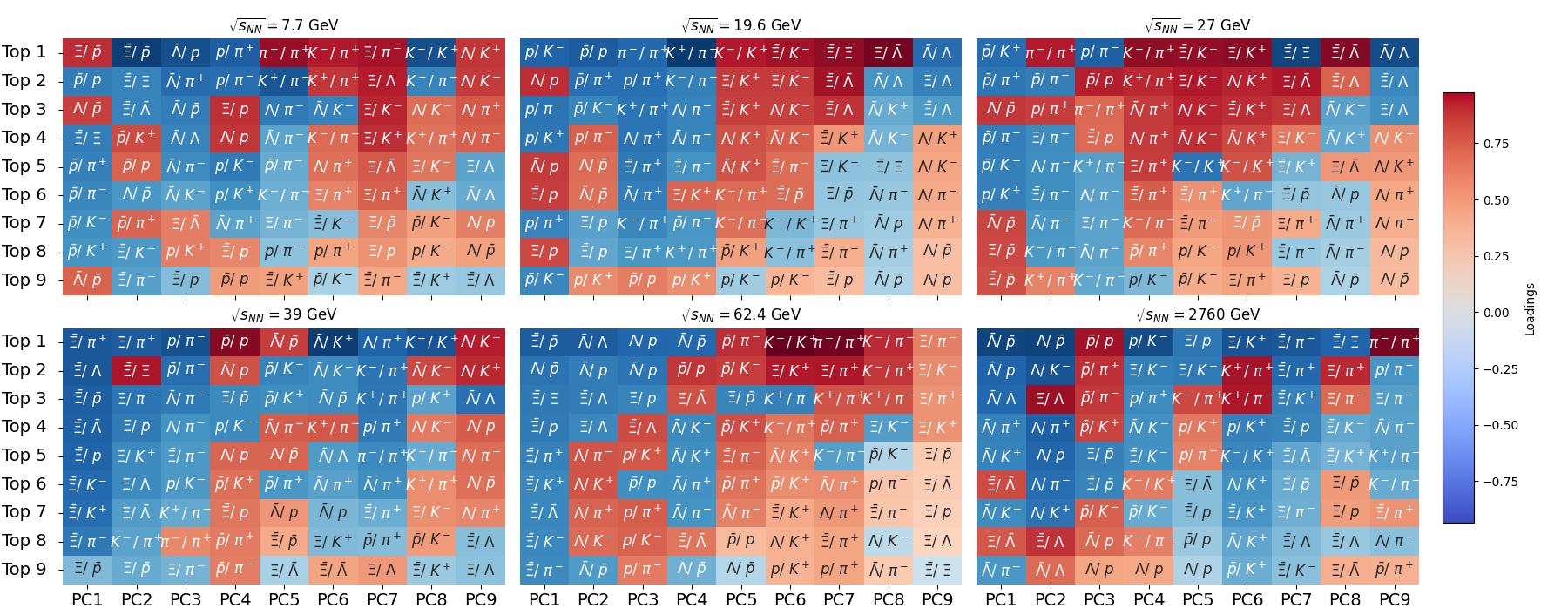}
		\caption{
			Energy dependence of the PCA loading matrices for hadron-yield log ratios.
			Each panel corresponds to a different collision energy, with the nine leading principal components (PCs) shown column-wise.
			For each PC, only the nine most significant and algebraically independent ratios with the largest loadings are displayed.
			The common color bar indicates the magnitude and sign of the loadings.
		}
		\label{fig:PCA_energy_wise}
	\end{figure*}

\begin{figure*}[t]
		\centering
		\includegraphics[scale=0.4]{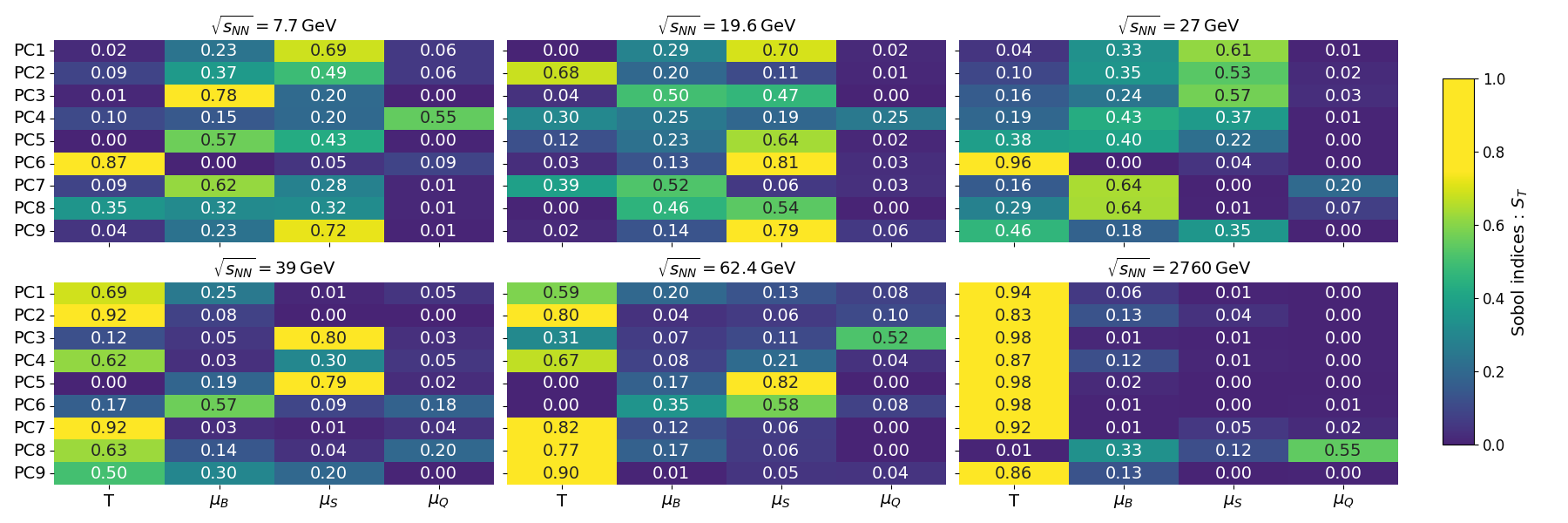}
		\caption{
			Sobol sensitivity analysis of the leading principal components (PCs) across collision energies.
			The color scale quantifies the fractional variance in each PC explained by the freeze-out parameters
			($T$, $\mu_B$, $\mu_S$, and $\mu_Q$), illustrating the transition from chemical-potential– to temperature-dominated freeze-out.
		}
		\label{fig:Sobol_PCA_plot}
	\end{figure*}

\begin{figure}[t]
	\centering
	\includegraphics[scale=0.3]{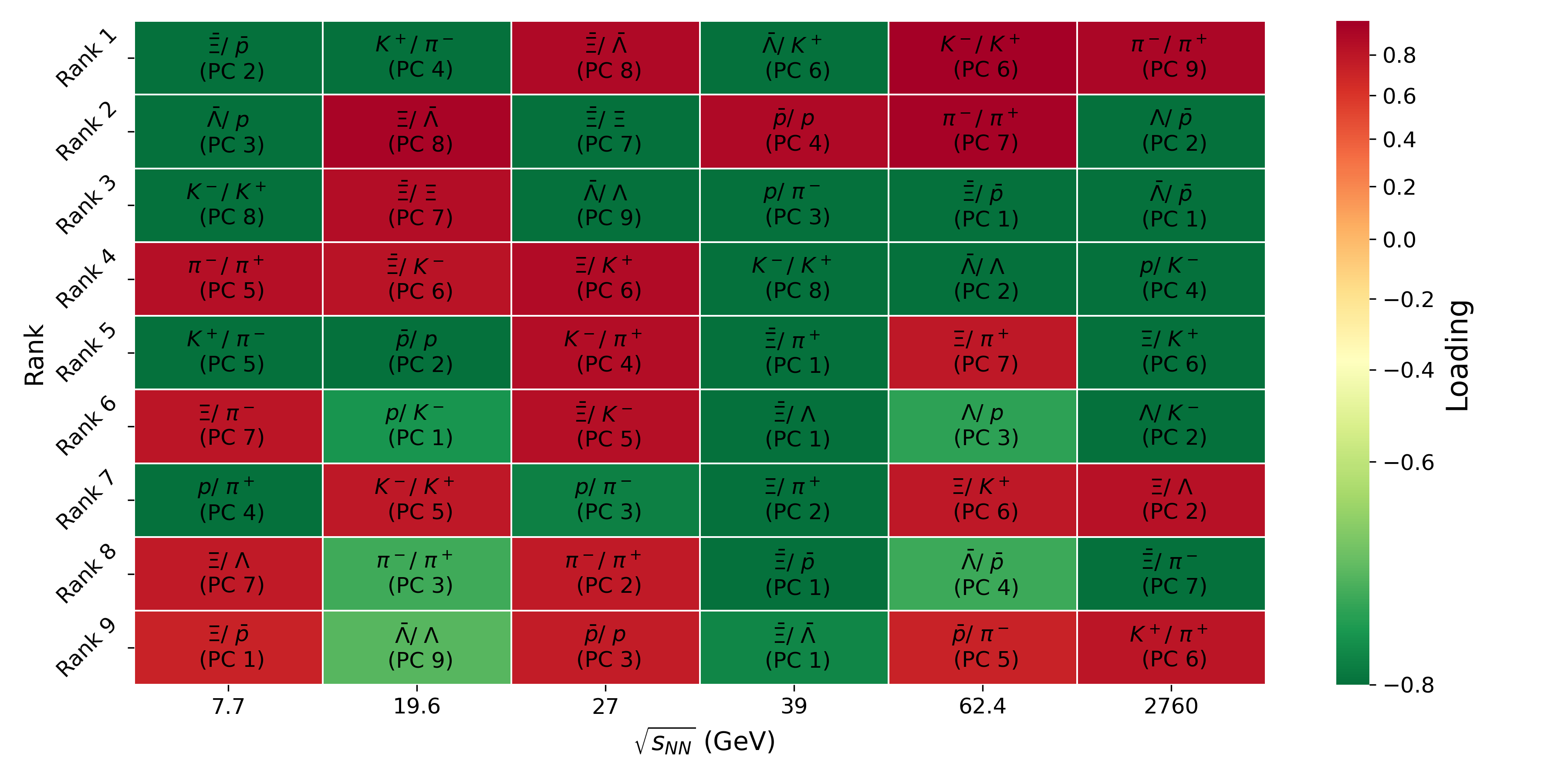}
	\caption{
		Energy dependence of the largest loadings across all principal components (PCs). 
		Each column corresponds to a different collision energy, showing the nine algebraically independent hadron-yield ratios with the largest absolute loadings across all PCs. 
		These ratios represent the most information-rich observables at each energy. 
	}
	\label{fig:loading_pca_energy_wise}
\end{figure}

	Figure~\ref{fig:PCA_energy_wise} shows, for each collision energy, the full PCA loading matrices, including all nine algebraically independent hadron-yield ratios with the largest absolute loadings across the principal components. In addition, the corresponding energy dependence of the Sobol total-effect indices for the principal-component amplitudes is presented in Fig.~\ref{fig:Sobol_PCA_plot}. For completeness and to aid interpretation, Fig.~\ref{fig:loading_pca_energy_wise} summarizes the nine algebraically independent ratios with the largest absolute loadings across all PCs at each collision energy. The detailed explanations of these figures are provided in their respective captions.

	\bibliographystyle{unsrt}  

	\bibliography{manuscript_5.bib}	

@article{Shuryak:1980tp,
    author = "Shuryak, Edward V.",
    title = "{Quantum Chromodynamics and the Theory of Superdense Matter}",
    doi = "10.1016/0370-1573(80)90105-2",
    journal = "Phys. Rept.",
    volume = "61",
    pages = "71--158",
    year = "1980"
}

@article{Braun-Munzinger:1994ewq,
    author = "Braun-Munzinger, P. and Stachel, J. and Wessels, J. P. and Xu, N.",
    title = "{Thermal equilibration and expansion in nucleus-nucleus collisions at the AGS}",
    eprint = "nucl-th/9410026",
    archivePrefix = "arXiv",
    reportNumber = "SUNY-RHI-94-11",
    doi = "10.1016/0370-2693(94)01534-J",
    journal = "Phys. Lett. B",
    volume = "344",
    pages = "43--48",
    year = "1995"
}

@article{Cleymans:1999st,
    author = "Cleymans, J. and Redlich, K.",
    title = "{Chemical and thermal freezeout parameters from 1-A/GeV to 200-A/GeV}",
    eprint = "nucl-th/9903063",
    archivePrefix = "arXiv",
    reportNumber = "BI-TP-99-19",
    doi = "10.1103/PhysRevC.60.054908",
    journal = "Phys. Rev. C",
    volume = "60",
    pages = "054908",
    year = "1999"
}

@article{Stephanov:1998dy,
    author = "Stephanov, Misha A. and Rajagopal, K. and Shuryak, Edward V.",
    title = "{Signatures of the tricritical point in QCD}",
    eprint = "hep-ph/9806219",
    archivePrefix = "arXiv",
    reportNumber = "ITP-SB-98-39, MIT-CTP-2748, SUNY-NTG-98-17",
    doi = "10.1103/PhysRevLett.81.4816",
    journal = "Phys. Rev. Lett.",
    volume = "81",
    pages = "4816--4819",
    year = "1998"
}

@article{Stephanov:1999zu,
    author = "Stephanov, Misha A. and Rajagopal, K. and Shuryak, Edward V.",
    title = "{Event-by-event fluctuations in heavy ion collisions and the QCD critical point}",
    eprint = "hep-ph/9903292",
    archivePrefix = "arXiv",
    reportNumber = "ITP-SB-99-4, MIT-CTP-2834, SUNY-NTG-99-3",
    doi = "10.1103/PhysRevD.60.114028",
    journal = "Phys. Rev. D",
    volume = "60",
    pages = "114028",
    year = "1999"
}

@article{Stephanov:2004wx,
    author = "Stephanov, Mikhail A.",
    editor = "Muller, Berndt and Tan, C. I.",
    title = "{QCD Phase Diagram and the Critical Point}",
    eprint = "hep-ph/0402115",
    archivePrefix = "arXiv",
    doi = "10.1143/PTPS.153.139",
    journal = "Prog. Theor. Phys. Suppl.",
    volume = "153",
    pages = "139--156",
    year = "2004"
}

@article{Tawfik:2004sw,
    author = "Tawfik, A.",
    title = "{QCD phase diagram: A Comparison of lattice and hadron resonance gas model calculations}",
    eprint = "hep-ph/0412336",
    archivePrefix = "arXiv",
    reportNumber = "BI-TH-2004-28",
    doi = "10.1103/PhysRevD.71.054502",
    journal = "Phys. Rev. D",
    volume = "71",
    pages = "054502",
    year = "2005"
}

@article{Cleymans:1996cd,
    author = "Cleymans, J. and Elliott, D. and Satz, H. and Thews, R. L.",
    title = "{Thermal hadron production in Si - Au collisions}",
    eprint = "nucl-th/9603004",
    archivePrefix = "arXiv",
    reportNumber = "CERN-TH-95-298, AZPH-TH-95-26, BI-TP-95-35, UCT-TP-95-225",
    doi = "10.1007/s002880050393",
    journal = "Z. Phys. C",
    volume = "74",
    pages = "319--324",
    year = "1997"
}

@article{Cleymans:1998fq,
    author = "Cleymans, J. and Redlich, K.",
    title = "{Unified description of freezeout parameters in relativistic heavy ion collisions}",
    eprint = "nucl-th/9808030",
    archivePrefix = "arXiv",
    doi = "10.1103/PhysRevLett.81.5284",
    journal = "Phys. Rev. Lett.",
    volume = "81",
    pages = "5284--5286",
    year = "1998"
}

@article{Braun-Munzinger:1999hun,
    author = "Braun-Munzinger, P. and Heppe, I. and Stachel, J.",
    title = "{Chemical equilibration in Pb + Pb collisions at the SPS}",
    eprint = "nucl-th/9903010",
    archivePrefix = "arXiv",
    doi = "10.1016/S0370-2693(99)01076-X",
    journal = "Phys. Lett. B",
    volume = "465",
    pages = "15--20",
    year = "1999"
}

@article{Braun-Munzinger:2001hwo,
    author = "Braun-Munzinger, P. and Magestro, D. and Redlich, K. and Stachel, J.",
    title = "{Hadron production in Au - Au collisions at RHIC}",
    eprint = "hep-ph/0105229",
    archivePrefix = "arXiv",
    reportNumber = "GSI-2001-15",
    doi = "10.1016/S0370-2693(01)01069-3",
    journal = "Phys. Lett. B",
    volume = "518",
    pages = "41--46",
    year = "2001"
}

@article{Becattini:2003wp,
    author = "Becattini, F. and Gazdzicki, M. and Keranen, A. and Manninen, J. and Stock, R.",
    title = "{Chemical equilibrium in nucleus nucleus collisions at relativistic energies}",
    eprint = "hep-ph/0310049",
    archivePrefix = "arXiv",
    doi = "10.1103/PhysRevC.69.024905",
    journal = "Phys. Rev. C",
    volume = "69",
    pages = "024905",
    year = "2004"
}

@article{Braun-Munzinger:2003pwq,
    author = "Braun-Munzinger, Peter and Redlich, Krzysztof and Stachel, Johanna",
    editor = "Hwa, Rudolph C. and Wang, Xin-Nian",
    title = "{Particle production in heavy ion collisions}",
    eprint = "nucl-th/0304013",
    archivePrefix = "arXiv",
    reportNumber = "GSI-PREPRINT-2003-13",
    doi = "10.1142/9789812795533_0008",
    pages = "491--599",
    month = "4",
    year = "2003"
}

@article{Andronic:2005yp,
    author = "Andronic, A. and Braun-Munzinger, P. and Stachel, J.",
    title = "{Hadron production in central nucleus-nucleus collisions at chemical freeze-out}",
    eprint = "nucl-th/0511071",
    archivePrefix = "arXiv",
    doi = "10.1016/j.nuclphysa.2006.03.012",
    journal = "Nucl. Phys. A",
    volume = "772",
    pages = "167--199",
    year = "2006"
}

@article{Becattini:2005xt,
    author = "Becattini, F. and Manninen, J. and Gazdzicki, M.",
    title = "{Energy and system size dependence of chemical freeze-out in relativistic nuclear collisions}",
    eprint = "hep-ph/0511092",
    archivePrefix = "arXiv",
    doi = "10.1103/PhysRevC.73.044905",
    journal = "Phys. Rev. C",
    volume = "73",
    pages = "044905",
    year = "2006"
}

@article{Andronic:2008gu,
    author = "Andronic, A. and Braun-Munzinger, P. and Stachel, J.",
    title = "{Thermal hadron production in relativistic nuclear collisions: The Hadron mass spectrum, the horn, and the QCD phase transition}",
    eprint = "0812.1186",
    archivePrefix = "arXiv",
    primaryClass = "nucl-th",
    doi = "10.1016/j.physletb.2009.06.021",
    journal = "Phys. Lett. B",
    volume = "673",
    pages = "142--145",
    year = "2009",
    note = "[Erratum: Phys.Lett.B 678, 516 (2009)]"
}

@article{Andronic:2011yq,
    author = "Andronic, A. and Braun-Munzinger, P. and Redlich, K. and Stachel, J.",
    editor = "Schutz, Yves and Wiedemann, Urs Achim",
    title = "{The thermal model on the verge of the ultimate test: particle production in Pb-Pb collisions at the LHC}",
    eprint = "1106.6321",
    archivePrefix = "arXiv",
    primaryClass = "nucl-th",
    doi = "10.1088/0954-3899/38/12/124081",
    journal = "J. Phys. G",
    volume = "38",
    pages = "124081",
    year = "2011"
}

@article{Begun:2012rf,
    author = "Begun, V. V. and Gazdzicki, M. and Gorenstein, M. I.",
    title = "{Hadron-resonance gas at freeze-out: Reminder on the importance of repulsive interactions}",
    eprint = "1208.4107",
    archivePrefix = "arXiv",
    primaryClass = "nucl-th",
    doi = "10.1103/PhysRevC.88.024902",
    journal = "Phys. Rev. C",
    volume = "88",
    number = "2",
    pages = "024902",
    year = "2013"
}

@article{Andronic:2017pug,
    author = "Andronic, Anton and Braun-Munzinger, Peter and Redlich, Krzysztof and Stachel, Johanna",
    title = "{Decoding the phase structure of QCD via particle production at high energy}",
    eprint = "1710.09425",
    archivePrefix = "arXiv",
    primaryClass = "nucl-th",
    doi = "10.1038/s41586-018-0491-6",
    journal = "Nature",
    volume = "561",
    number = "7723",
    pages = "321--330",
    year = "2018"
}

@article{Chatterjee:2017yhp,
    author = "Chatterjee, Sandeep and Mishra, Debadeepti and Mohanty, Bedangadas and Samanta, Subhasis",
    title = "{Freezeout systematics due to the hadron spectrum}",
    eprint = "1708.08152",
    archivePrefix = "arXiv",
    primaryClass = "nucl-th",
    doi = "10.1103/PhysRevC.96.054907",
    journal = "Phys. Rev. C",
    volume = "96",
    number = "5",
    pages = "054907",
    year = "2017"
}

@article{Alba:2014eba,
    author = "Alba, Paolo and Alberico, Wanda and Bellwied, Rene and Bluhm, Marcus and Mantovani Sarti, Valentina and Nahrgang, Marlene and Ratti, Claudia",
    title = "{Freeze-out conditions from net-proton and net-charge fluctuations at RHIC}",
    eprint = "1403.4903",
    archivePrefix = "arXiv",
    primaryClass = "hep-ph",
    doi = "10.1016/j.physletb.2014.09.052",
    journal = "Phys. Lett. B",
    volume = "738",
    pages = "305--310",
    year = "2014"
}

@article{Adak:2016jtk,
    author = "Adak, Rama Prasad and Das, Supriya and Ghosh, Sanjay K. and Ray, Rajarshi and Samanta, Subhasis",
    title = "{Centrality dependence of chemical freeze-out parameters from net-proton and net-charge fluctuations using a hadron resonance gas model}",
    eprint = "1609.05318",
    archivePrefix = "arXiv",
    primaryClass = "nucl-th",
    doi = "10.1103/PhysRevC.96.014902",
    journal = "Phys. Rev. C",
    volume = "96",
    number = "1",
    pages = "014902",
    year = "2017"
}

@article{Cleymans:2005xv,
    author = "Cleymans, J. and Oeschler, H. and Redlich, K. and Wheaton, S.",
    title = "{Comparison of chemical freeze-out criteria in heavy-ion collisions}",
    eprint = "hep-ph/0511094",
    archivePrefix = "arXiv",
    reportNumber = "CERN-PH-TH-2005-210, CERN-PH-TH/2005-210",
    doi = "10.1103/PhysRevC.73.034905",
    journal = "Phys. Rev. C",
    volume = "73",
    pages = "034905",
    year = "2006"
}

@article{Chatterjee:2013yga,
    author = "Chatterjee, S. and Godbole, R. M. and Gupta, Sourendu",
    title = "{Strange freezeout}",
    eprint = "1306.2006",
    archivePrefix = "arXiv",
    primaryClass = "nucl-th",
    reportNumber = "TIFR-TH-13-15",
    doi = "10.1016/j.physletb.2013.11.008",
    journal = "Phys. Lett. B",
    volume = "727",
    pages = "554--557",
    year = "2013"
}

@article{Vovchenko:2015cbk,
    author = {Vovchenko, Volodymyr and St{\"o}cker, Horst},
    title = "{Surprisingly large uncertainties in temperature extraction from thermal fits to hadron yield data at LHC}",
    eprint = "1512.08046",
    archivePrefix = "arXiv",
    primaryClass = "hep-ph",
    doi = "10.1088/1361-6471/aa6409",
    journal = "J. Phys. G",
    volume = "44",
    number = "5",
    pages = "055103",
    year = "2017"
}

@article{Vovchenko:2016ebv,
    author = "Vovchenko, Volodymyr and Stoecker, Horst",
    title = "{Examination of the sensitivity of the thermal fits to heavy-ion hadron yield data to the modeling of the eigenvolume interactions}",
    eprint = "1606.06218",
    archivePrefix = "arXiv",
    primaryClass = "hep-ph",
    doi = "10.1103/PhysRevC.95.044904",
    journal = "Phys. Rev. C",
    volume = "95",
    number = "4",
    pages = "044904",
    year = "2017"
}

@article{Alba:2016hwx,
    author = "Alba, P. and Vovchenko, V. and Gorenstein, M. I. and Stoecker, H.",
    title = "{Flavor-dependent eigenvolume interactions in a hadron resonance gas}",
    eprint = "1606.06542",
    archivePrefix = "arXiv",
    primaryClass = "hep-ph",
    doi = "10.1016/j.nuclphysa.2018.03.007",
    journal = "Nucl. Phys. A",
    volume = "974",
    pages = "22--34",
    year = "2018"
}

@article{Alba:2017bbr,
    author = "Alba, Paolo and Oliva, L.",
    title = "{Balance of attractive and repulsive hadronic interactions: The influence of hadronic spectrum and excluded-volume effects on lattice thermodynamics, and consequences for experiments}",
    eprint = "1711.02797",
    archivePrefix = "arXiv",
    primaryClass = "nucl-th",
    doi = "10.1103/PhysRevC.99.055207",
    journal = "Phys. Rev. C",
    volume = "99",
    number = "5",
    pages = "055207",
    year = "2019"
}

@article{Sarkar:2025bkc,
    author = "Sarkar, Nachiketa",
    title = "{Bayesian Calibration of the Crossterms Eigenvolume HRG Model: Integrating Lattice QCD and Experimental Data}",
    eprint = "2508.12266",
    archivePrefix = "arXiv",
    primaryClass = "hep-ph",
    month = "8",
    year = "2025"
}

@article{Yen:1997rv,
    author = "Yen, Granddon D. and Gorenstein, Mark I. and Greiner, Walter and Yang, Shin-Nan",
    title = "{Excluded volume hadron gas model for particle number ratios in A+A collisions}",
    eprint = "nucl-th/9711062",
    archivePrefix = "arXiv",
    doi = "10.1103/PhysRevC.56.2210",
    journal = "Phys. Rev. C",
    volume = "56",
    pages = "2210--2218",
    year = "1997"
}

@article{Tiwari:2011km,
    author = "Tiwari, S. K. and Srivastava, P. K. and Singh, C. P.",
    title = "{Description of Hot and Dense Hadron Gas Properties in a New Excluded-Volume model}",
    eprint = "1111.2406",
    archivePrefix = "arXiv",
    primaryClass = "hep-ph",
    doi = "10.1103/PhysRevC.85.014908",
    journal = "Phys. Rev. C",
    volume = "85",
    pages = "014908",
    year = "2012"
}

@article{Rischke:1991ke,
    author = "Rischke, Dirk H. and Gorenstein, Mark I. and Stoecker, Horst and Greiner, Walter",
    title = "{Excluded volume effect for the nuclear matter equation of state}",
    reportNumber = "UFTP-252-1990",
    doi = "10.1007/BF01548574",
    journal = "Z. Phys. C",
    volume = "51",
    pages = "485--490",
    year = "1991"
}

@article{Cleymans:1992jz,
    author = "Cleymans, J. and Gorenstein, Mark I. and Stalnacke, J. and Suhonen, E.",
    title = "{Excluded volume effect and the quark - hadron phase transition}",
    reportNumber = "BI-TP-92-16",
    doi = "10.1088/0031-8949/48/3/004",
    journal = "Phys. Scripta",
    volume = "48",
    pages = "277--280",
    year = "1993"
}

@article{Andronic:2012ut,
    author = "Andronic, A. and Braun-Munzinger, P. and Stachel, J. and Winn, M.",
    title = "{Interacting hadron resonance gas meets lattice QCD}",
    eprint = "1201.0693",
    archivePrefix = "arXiv",
    primaryClass = "nucl-th",
    doi = "10.1016/j.physletb.2012.10.001",
    journal = "Phys. Lett. B",
    volume = "718",
    pages = "80--85",
    year = "2012"
}

@article{Bhattacharyya:2013oya,
    author = "Bhattacharyya, Abhijit and Das, Supriya and Ghosh, Sanjay K. and Ray, Rajarshi and Samanta, Subhasis",
    title = "{Fluctuations and correlations of conserved charges in an excluded volume hadron resonance gas model}",
    eprint = "1310.2793",
    archivePrefix = "arXiv",
    primaryClass = "hep-ph",
    doi = "10.1103/PhysRevC.90.034909",
    journal = "Phys. Rev. C",
    volume = "90",
    number = "3",
    pages = "034909",
    year = "2014"
}

@article{Vovchenko:2016rkn,
    author = "Vovchenko, Volodymyr and Gorenstein, Mark I. and Stoecker, Horst",
    title = "{van der Waals Interactions in Hadron Resonance Gas: From Nuclear Matter to Lattice QCD}",
    eprint = "1609.03975",
    archivePrefix = "arXiv",
    primaryClass = "hep-ph",
    doi = "10.1103/PhysRevLett.118.182301",
    journal = "Phys. Rev. Lett.",
    volume = "118",
    number = "18",
    pages = "182301",
    year = "2017"
}

@article{Sarkar:2018mbk,
    author = "Sarkar, Nachiketa and Ghosh, Premomoy",
    title = "{van der Waals hadron resonance gas and QCD phase diagram}",
    eprint = "1807.02948",
    archivePrefix = "arXiv",
    primaryClass = "hep-ph",
    doi = "10.1103/PhysRevC.98.014907",
    journal = "Phys. Rev. C",
    volume = "98",
    number = "1",
    pages = "014907",
    year = "2018"
}

@article{Vovchenko:2014pka,
    author = "Vovchenko, V. and Anchishkin, D. V. and Gorenstein, M. I.",
    title = "{Hadron Resonance Gas Equation of State from Lattice QCD}",
    eprint = "1412.5478",
    archivePrefix = "arXiv",
    primaryClass = "nucl-th",
    doi = "10.1103/PhysRevC.91.024905",
    journal = "Phys. Rev. C",
    volume = "91",
    number = "2",
    pages = "024905",
    year = "2015"
}

@article{Sarkar:2017ijd,
    author = "Sarkar, Nachiketa and Ghosh, Premomoy",
    title = "{Thermalization in a small hadron gas system and high-multiplicity $pp$ events}",
    eprint = "1706.08679",
    archivePrefix = "arXiv",
    primaryClass = "hep-ph",
    doi = "10.1103/PhysRevC.96.044901",
    journal = "Phys. Rev. C",
    volume = "96",
    number = "4",
    pages = "044901",
    year = "2017"
}

@article{Karthein:2021cmb,
    author = "Karthein, Jamie M. and Koch, Volker and Ratti, Claudia and Vovchenko, Volodymyr",
    title = "{Constraining the hadronic spectrum and repulsive interactions in a hadron resonance gas via fluctuations of conserved charges}",
    eprint = "2107.00588",
    archivePrefix = "arXiv",
    primaryClass = "nucl-th",
    doi = "10.1103/PhysRevD.104.094009",
    journal = "Phys. Rev. D",
    volume = "104",
    number = "9",
    pages = "094009",
    year = "2021"
}

@article{Noronha-Hostler:2012ycm,
    author = "Noronha-Hostler, Jacquelyn and Noronha, Jorge and Greiner, Carsten",
    title = "{Hadron Mass Spectrum and the Shear Viscosity to Entropy Density Ratio of Hot Hadronic Matter}",
    eprint = "1206.5138",
    archivePrefix = "arXiv",
    primaryClass = "nucl-th",
    doi = "10.1103/PhysRevC.86.024913",
    journal = "Phys. Rev. C",
    volume = "86",
    pages = "024913",
    year = "2012"
}

@article{Sarkar:2017bqy,
    author = "Sarkar, Nachiketa and Ghosh, Premomoy",
    title = "{The $\eta$/s of the LQCD-EoS complied hadron gas of different sizes approach common minimum near the crossover temperature}",
    eprint = "1711.08721",
    archivePrefix = "arXiv",
    primaryClass = "hep-ph",
    doi = "10.1140/epja/i2018-12623-2",
    journal = "Eur. Phys. J. A",
    volume = "54",
    number = "11",
    pages = "194",
    year = "2018"
}

@article{Albright:2014gva,
    author = "Albright, M. and Kapusta, J. and Young, C.",
    title = "{Matching Excluded Volume Hadron Resonance Gas Models and Perturbative QCD to Lattice Calculations}",
    eprint = "1404.7540",
    archivePrefix = "arXiv",
    primaryClass = "nucl-th",
    doi = "10.1103/PhysRevC.90.024915",
    journal = "Phys. Rev. C",
    volume = "90",
    number = "2",
    pages = "024915",
    year = "2014"
}

@article{Sarkar:2019oyo,
    author = "Sarkar, Nachiketa and Deb, Paramita and Ghosh, Premomoy",
    title = "{Finite size effect on thermodynamics of hadron gas in high-multiplicity events of proton-proton collisions at the LHC}",
    eprint = "1905.06532",
    archivePrefix = "arXiv",
    primaryClass = "hep-ph",
    month = "5",
    year = "2019"
}

@article{Venugopalan:1992hy,
    author = "Venugopalan, R. and Prakash, M.",
    title = "{Thermal properties of interacting hadrons}",
    doi = "10.1016/0375-9474(92)90005-5",
    journal = "Nucl. Phys. A",
    volume = "546",
    pages = "718--760",
    year = "1992"
}

@article{Dash:2018can,
    author = "Dash, Ashutosh and Samanta, Subhasis and Mohanty, Bedangadas",
    title = "{Interacting hadron resonance gas model in the K -matrix formalism}",
    eprint = "1802.04998",
    archivePrefix = "arXiv",
    primaryClass = "nucl-th",
    doi = "10.1103/PhysRevC.97.055208",
    journal = "Phys. Rev. C",
    volume = "97",
    number = "5",
    pages = "055208",
    year = "2018"
}

@article{Vovchenko:2017drx,
    author = "Vovchenko, Volodymyr and Motornenko, Anton and Gorenstein, Mark I. and Stoecker, Horst",
    title = "{Beth-Uhlenbeck approach for repulsive interactions between baryons in a hadron gas}",
    eprint = "1710.00693",
    archivePrefix = "arXiv",
    primaryClass = "nucl-th",
    doi = "10.1103/PhysRevC.97.035202",
    journal = "Phys. Rev. C",
    volume = "97",
    number = "3",
    pages = "035202",
    year = "2018"
}

@article{Dash:2018mep,
    author = "Dash, Ashutosh and Samanta, Subhasis and Mohanty, Bedangadas",
    title = "{Thermodynamics of a gas of hadrons with attractive and repulsive interactions within an S -matrix formalism}",
    eprint = "1806.02117",
    archivePrefix = "arXiv",
    primaryClass = "hep-ph",
    doi = "10.1103/PhysRevC.99.044919",
    journal = "Phys. Rev. C",
    volume = "99",
    number = "4",
    pages = "044919",
    year = "2019"
}

@article{Koch:1986ud,
    author = "Koch, P. and Muller, Berndt and Rafelski, Johann",
    title = "{Strangeness in Relativistic Heavy Ion Collisions}",
    reportNumber = "GSI-86-7, UCT-TP-41-86, UCT-TP-29-1985",
    doi = "10.1016/0370-1573(86)90096-7",
    journal = "Phys. Rept.",
    volume = "142",
    pages = "167--262",
    year = "1986"
}

@article{Manninen:2008mg,
    author = "Manninen, J. and Becattini, F.",
    title = "{Chemical freeze-out in ultra-relativistic heavy ion collisions at s(NN)**(1/2) = 130 and 200-GeV}",
    eprint = "0806.4100",
    archivePrefix = "arXiv",
    primaryClass = "nucl-th",
    doi = "10.1103/PhysRevC.78.054901",
    journal = "Phys. Rev. C",
    volume = "78",
    pages = "054901",
    year = "2008"
}

@article{Bugaev:2013sfa,
    author = "Bugaev, K. A. and Oliinychenko, D. R. and Cleymans, J. and Ivanytskyi, A. I. and Mishustin, I. N. and Nikonov, E. G. and Sagun, V. V.",
    title = "{Chemical Freeze-out of Strange Particles and Possible Root of Strangeness Suppression}",
    eprint = "1308.3594",
    archivePrefix = "arXiv",
    primaryClass = "hep-ph",
    doi = "10.1209/0295-5075/104/22002",
    journal = "EPL",
    volume = "104",
    number = "2",
    pages = "22002",
    year = "2013"
}

@article{Chatterjee:2014ysa,
    author = "Chatterjee, Sandeep and Mohanty, Bedangadas",
    title = "{Production of Light Nuclei in Heavy Ion Collisions Within Multiple Freezeout Scenario}",
    eprint = "1405.2632",
    archivePrefix = "arXiv",
    primaryClass = "nucl-th",
    doi = "10.1103/PhysRevC.90.034908",
    journal = "Phys. Rev. C",
    volume = "90",
    number = "3",
    pages = "034908",
    year = "2014"
}

@article{Flor:2020fdw,
    author = "Flor, Fernando Antonio and Olinger, Gabrielle and Bellwied, Rene",
    title = "{Flavour and Energy Dependence of Chemical Freeze-out Temperatures in Relativistic Heavy Ion Collisions from RHIC-BES to LHC Energies}",
    eprint = "2009.14781",
    archivePrefix = "arXiv",
    primaryClass = "nucl-ex",
    doi = "10.1016/j.physletb.2021.136098",
    journal = "Phys. Lett. B",
    volume = "814",
    pages = "136098",
    year = "2021"
}

@article{Sharma:2018jqf,
    author = "Sharma, Natasha and Cleymans, Jean and Hippolyte, Boris and Paradza, Masimba",
    title = "{A Comparison of p-p, p-Pb, Pb-Pb Collisions in the Thermal Model: Multiplicity Dependence of Thermal Parameters}",
    eprint = "1811.00399",
    archivePrefix = "arXiv",
    primaryClass = "hep-ph",
    doi = "10.1103/PhysRevC.99.044914",
    journal = "Phys. Rev. C",
    volume = "99",
    number = "4",
    pages = "044914",
    year = "2019"
}

@article{Panda:2021zab,
    author = "Panda, Susil Kumar and Chatterjee, Sandeep and Dash, Ajay Kumar and Mohanty, Bedangadas and Paikaray, Rita and Samanta, Subhasis and Singh, Ranbir",
    title = "{Multiplicity dependence of freezeout scenarios in pp collisions at s=7 TeV}",
    eprint = "2112.04226",
    archivePrefix = "arXiv",
    primaryClass = "hep-ph",
    doi = "10.1103/PhysRevC.104.064905",
    journal = "Phys. Rev. C",
    volume = "104",
    number = "6",
    pages = "064905",
    year = "2021"
}

@article{Flor:2021olm,
    author = "Flor, Fernando Antonio and Olinger, Gabrielle and Bellwied, Ren{\'e}",
    title = "{System size and flavour dependence of chemical freeze-out temperatures in ALICE data from pp, pPb and PbPb collisions at LHC energies}",
    eprint = "2109.09843",
    archivePrefix = "arXiv",
    primaryClass = "nucl-ex",
    doi = "10.1016/j.physletb.2022.137473",
    journal = "Phys. Lett. B",
    volume = "834",
    pages = "137473",
    year = "2022"
}

@article{Bhattacharyya:2019cer,
    author = "Bhattacharyya, Sumana and Biswas, Deeptak and Ghosh, Sanjay K. and Ray, Rajarshi and Singha, Pracheta",
    title = "{Systematics of chemical freeze-out parameters in heavy-ion collision experiments}",
    eprint = "1911.04828",
    archivePrefix = "arXiv",
    primaryClass = "hep-ph",
    doi = "10.1103/PhysRevD.101.054002",
    journal = "Phys. Rev. D",
    volume = "101",
    number = "5",
    pages = "054002",
    year = "2020"
}

@article{Bhattacharyya:2020sgn,
    author = "Bhattacharyya, Sumana and Jaiswal, Amaresh and Roy, Sutanu",
    title = "{Chemical freeze-out systematics of thermal model analysis using hadron yield ratios}",
    eprint = "2009.13399",
    archivePrefix = "arXiv",
    primaryClass = "hep-ph",
    doi = "10.1103/PhysRevC.103.024905",
    journal = "Phys. Rev. C",
    volume = "103",
    number = "2",
    pages = "024905",
    year = "2021"
}

@article{Becattini:2007wt,
    author = "Becattini, F.",
    title = "{Remark on statistical model fits to particle ratios in relativistic heavy ion collisions}",
    eprint = "0707.4154",
    archivePrefix = "arXiv",
    primaryClass = "nucl-th",
    month = "7",
    year = "2007"
}

@article{Torrieri:2004zz,
    author = "Torrieri, Giorgio and Steinke, S. and Broniowski, Wojciech and Florkowski, Wojciech and Letessier, Jean and Rafelski, Johann",
    title = "{SHARE: Statistical hadronization with resonances}",
    eprint = "nucl-th/0404083",
    archivePrefix = "arXiv",
    doi = "10.1016/j.cpc.2005.01.004",
    journal = "Comput. Phys. Commun.",
    volume = "167",
    pages = "229--251",
    year = "2005"
}

@article{Bhattacharyya:2019wag,
    author = "Bhattacharyya, Sumana and Biswas, Deeptak and Ghosh, Sanjay K. and Ray, Rajarshi and Singha, Pracheta",
    title = "{Novel scheme for parametrizing the chemical freeze-out surface in Heavy Ion Collision Experiments}",
    eprint = "1904.00959",
    archivePrefix = "arXiv",
    primaryClass = "nucl-th",
    doi = "10.1103/PhysRevD.100.054037",
    journal = "Phys. Rev. D",
    volume = "100",
    number = "5",
    pages = "054037",
    year = "2019"
}

@article{Karsch:2003zq,
    author = "Karsch, F. and Redlich, K. and Tawfik, A.",
    title = "{Thermodynamics at nonzero baryon number density: A Comparison of lattice and hadron resonance gas model calculations}",
    eprint = "hep-ph/0306208",
    archivePrefix = "arXiv",
    reportNumber = "BI-TP-2003-16",
    doi = "10.1016/j.physletb.2003.08.001",
    journal = "Phys. Lett. B",
    volume = "571",
    pages = "67--74",
    year = "2003"
}

@article{Garg:2013ata,
    author = "Garg, P. and Mishra, D. K. and Netrakanti, P. K. and Mohanty, B. and Mohanty, A. K. and Singh, B. K. and Xu, N.",
    title = "{Conserved number fluctuations in a hadron resonance gas model}",
    eprint = "1304.7133",
    archivePrefix = "arXiv",
    primaryClass = "nucl-ex",
    doi = "10.1016/j.physletb.2013.09.019",
    journal = "Phys. Lett. B",
    volume = "726",
    pages = "691--696",
    year = "2013"
}

@article{ParticleDataGroup:2020ssz,
    author = "Zyla, P. A. and others",
    collaboration = "Particle Data Group",
    title = "{Review of Particle Physics}",
    doi = "10.1093/ptep/ptaa104",
    journal = "PTEP",
    volume = "2020",
    number = "8",
    pages = "083C01",
    year = "2020"
}

@article{Bernhard:2016tnd,
    author = "Bernhard, Jonah E. and Moreland, J. Scott and Bass, Steffen A. and Liu, Jia and Heinz, Ulrich",
    title = "{Applying Bayesian parameter estimation to relativistic heavy-ion collisions: simultaneous characterization of the initial state and quark-gluon plasma medium}",
    eprint = "1605.03954",
    archivePrefix = "arXiv",
    primaryClass = "nucl-th",
    doi = "10.1103/PhysRevC.94.024907",
    journal = "Phys. Rev. C",
    volume = "94",
    number = "2",
    pages = "024907",
    year = "2016"
}

@article{JETSCAPE:2021ehl,
    author = "Cao, S. and others",
    collaboration = "JETSCAPE",
    title = "{Determining the jet transport coefficient $\cap{q}$ from inclusive hadron suppression measurements using Bayesian parameter estimation}",
    eprint = "2102.11337",
    archivePrefix = "arXiv",
    primaryClass = "nucl-th",
    doi = "10.1103/PhysRevC.104.024905",
    journal = "Phys. Rev. C",
    volume = "104",
    number = "2",
    pages = "024905",
    year = "2021"
}

@article{Bernhard:2019bmu,
    author = "Bernhard, Jonah E. and Moreland, J. Scott and Bass, Steffen A.",
    title = "{Bayesian estimation of the specific shear and bulk viscosity of quark{\textendash}gluon plasma}",
    doi = "10.1038/s41567-019-0611-8",
    journal = "Nature Phys.",
    volume = "15",
    number = "11",
    pages = "1113--1117",
    year = "2019"
}

@article{Wesolowski:2015fqa,
    author = "Wesolowski, S. and Klco, N. and Furnstahl, R. J. and Phillips, D. R. and Thapaliya, A.",
    title = "{Bayesian parameter estimation for effective field theories}",
    eprint = "1511.03618",
    archivePrefix = "arXiv",
    primaryClass = "nucl-th",
    doi = "10.1088/0954-3899/43/7/074001",
    journal = "J. Phys. G",
    volume = "43",
    number = "7",
    pages = "074001",
    year = "2016"
}

@article{Nijs:2020roc,
    author = {Nijs, Govert and van der Schee, Wilke and G{\"u}rsoy, Umut and Snellings, Raimond},
    title = "{Bayesian analysis of heavy ion collisions with the heavy ion computational framework Trajectum}",
    eprint = "2010.15134",
    archivePrefix = "arXiv",
    primaryClass = "nucl-th",
    reportNumber = "CERN-TH-2020-175, MIT-CTP/5251",
    doi = "10.1103/PhysRevC.103.054909",
    journal = "Phys. Rev. C",
    volume = "103",
    number = "5",
    pages = "054909",
    year = "2021"
}

@article{Parkkila:2021yha,
    author = "Parkkila, J. E. and Onnerstad, A. and Taghavi, S. F. and Mordasini, C. and Bilandzic, A. and Virta, M. and Kim, D. J.",
    title = "{New constraints for QCD matter from improved Bayesian parameter estimation in heavy-ion collisions at LHC}",
    eprint = "2111.08145",
    archivePrefix = "arXiv",
    primaryClass = "hep-ph",
    doi = "10.1016/j.physletb.2022.137485",
    journal = "Phys. Lett. B",
    volume = "835",
    pages = "137485",
    year = "2022"
}

@phdthesis{Bernhard:2018hnz,
    author = "Bernhard, Jonah E.",
    title = "{Bayesian parameter estimation for relativistic heavy-ion collisions}",
    eprint = "1804.06469",
    archivePrefix = "arXiv",
    primaryClass = "nucl-th",
    school = "Duke U.",
    month = "4",
    year = "2018"
}

@article{STAR:2017sal,
    author = "Adamczyk, L. and others",
    collaboration = "STAR",
    title = "{Bulk Properties of the Medium Produced in Relativistic Heavy-Ion Collisions from the Beam Energy Scan Program}",
    eprint = "1701.07065",
    archivePrefix = "arXiv",
    primaryClass = "nucl-ex",
    doi = "10.1103/PhysRevC.96.044904",
    journal = "Phys. Rev. C",
    volume = "96",
    number = "4",
    pages = "044904",
    year = "2017"
}

@article{ALICE:2013mez,
    author = "Abelev, Betty and others",
    collaboration = "ALICE",
    title = "{Centrality dependence of $\pi$, K, p production in Pb-Pb collisions at $\sqrt{s_{NN}}$ = 2.76 TeV}",
    eprint = "1303.0737",
    archivePrefix = "arXiv",
    primaryClass = "hep-ex",
    reportNumber = "CERN-PH-EP-2013-019",
    doi = "10.1103/PhysRevC.88.044910",
    journal = "Phys. Rev. C",
    volume = "88",
    pages = "044910",
    year = "2013"
}

@article{ALICE:2013cdo,
    author = "Abelev, Betty Bezverkhny and others",
    collaboration = "ALICE",
    title = "{$K^0_S$ and $\Lambda$ production in Pb-Pb collisions at $\sqrt{s_{NN}}$ = 2.76 TeV}",
    eprint = "1307.5530",
    archivePrefix = "arXiv",
    primaryClass = "nucl-ex",
    reportNumber = "CERN-PH-EP-2013-132",
    doi = "10.1103/PhysRevLett.111.222301",
    journal = "Phys. Rev. Lett.",
    volume = "111",
    pages = "222301",
    year = "2013"
}

@article{ALICE:2013xmt,
    author = "Abelev, Betty Bezverkhny and others",
    collaboration = "ALICE",
    title = "{Multi-strange baryon production at mid-rapidity in Pb-Pb collisions at $\sqrt{s_{NN}}$ = 2.76 TeV}",
    eprint = "1307.5543",
    archivePrefix = "arXiv",
    primaryClass = "nucl-ex",
    reportNumber = "CERN-PH-EP-2013-134",
    doi = "10.1016/j.physletb.2014.05.052",
    journal = "Phys. Lett. B",
    volume = "728",
    pages = "216--227",
    year = "2014",
    note = "[Erratum: Phys.Lett.B 734, 409--410 (2014)]"
}

@article{Becattini:2014hla,
    author = "Becattini, F. and Grossi, Eduardo and Bleicher, Marcus and Steinheimer, Jan and Stock, Reinhard",
    title = "{Centrality dependence of hadronization and chemical freeze-out conditions in heavy ion collisions at $\sqrt s_{NN}$ = 2.76 TeV}",
    eprint = "1405.0710",
    archivePrefix = "arXiv",
    primaryClass = "nucl-th",
    doi = "10.1103/PhysRevC.90.054907",
    journal = "Phys. Rev. C",
    volume = "90",
    number = "5",
    pages = "054907",
    year = "2014"
}

@article{Vovchenko:2015idt,
    author = "Vovchenko, V. and Begun, V. V. and Gorenstein, M. I.",
    title = "{Hadron multiplicities and chemical freeze-out conditions in proton-proton and nucleus-nucleus collisions}",
    eprint = "1512.08025",
    archivePrefix = "arXiv",
    primaryClass = "nucl-th",
    doi = "10.1103/PhysRevC.93.064906",
    journal = "Phys. Rev. C",
    volume = "93",
    number = "6",
    pages = "064906",
    year = "2016"
}

@article{Jolliffe2016,
  author    = {I. T. Jolliffe and J. Cadima},
  title     = {Principal Component Analysis: A Review and Recent Developments},
  journal   = {Philosophical Transactions of the Royal Society A: Mathematical, Physical and Engineering Sciences},
  volume    = {374},
  number    = {2065},
  pages     = {20150202},
  year      = {2016},
  doi       = {10.1098/rsta.2015.0202}
}

@article{Pearson1901,
  author    = {Karl Pearson},
  title     = {On Lines and Planes of Closest Fit to Systems of Points in Space},
  journal   = {Philosophical Magazine},
  volume    = {2},
  number    = {11},
  pages     = {559--572},
  year      = {1901},
  doi       = {10.1080/14786440109462720}
}

@article{Foreman-Mackey_2013,
doi = {10.1086/670067},
url = {https://doi.org/10.1086/670067},
year = {2013},
month = {feb},
publisher = {University of Chicago Press},
volume = {125},
number = {925},
pages = {306},
author = {Foreman-Mackey, Daniel and Hogg, David W. and Lang, Dustin and Goodman, Jonathan},
title = {emcee: The MCMC Hammer},
journal = {Publications of the Astronomical Society of the Pacific},
 }

@article{McKay1979LHS,
  title     = {A Comparison of Three Methods for Selecting Values of Input Variables in the Analysis of Output from a Computer Code},
  author    = {McKay, M. D. and Beckman, R. J. and Conover, W. J.},
  journal   = {Technometrics},
  volume    = {21},
  number    = {2},
  pages     = {239--245},
  year      = {1979},
  publisher = {Taylor \& Francis},
  doi       = {10.1080/00401706.1979.10489755}
}

@article{Sobol1993,
  author = {Sobol, I. M.},
  title = {Sensitivity estimates for nonlinear mathematical models},
  journal = {Mathematical Modelling and Computational Experiments},
  volume = {1},
  pages = {407--414},
  year = {1993}
}

@article{Saltelli2002,
  author = {Saltelli, A.},
  title = {Making best use of model evaluations to compute sensitivity indices},
  journal = {Computer Physics Communications},
  volume = {145},
  pages = {280--297},
  year = {2002}
}

@misc{HepBayes,
  author       = {Sarkar, Nachiketa},
  title        = {HepBayes: Bayesian Inference Framework for Heavy-Ion Collisions},
  year         = {2025},
  howpublished = {\url{https://github.com/nachiketa-sarkar/HepBayes}}
}

@article{Biswas:2020dsc,
    author = "Biswas, Deeptak",
    title = "{Centrality dependence of chemical freeze-out parameters and strangeness equilibration in RHIC and LHC energies}",
    eprint = "2003.10425",
    archivePrefix = "arXiv",
    primaryClass = "hep-ph",
    doi = "10.1155/2021/6611394",
    journal = "Adv. High Energy Phys.",
    volume = "2021",
    pages = "6611394",
    year = "2021"
}
	
\end{document}